\documentclass[twocolumn]{aastex631}

\usepackage{amsmath}
\usepackage{hhline}
\usepackage{url}
\usepackage{ulem}
\usepackage{soul}

\usepackage[caption=false]{subfig}
\usepackage{xcolor}

\begin{document}

\title{Machine learning based Photometric Redshifts for Galaxies in the North Ecliptic Pole Wide field: catalogs of spectroscopic and photometric redshifts}

\correspondingauthor{Ho Seong Hwang}
\email{hhwang@astro.snu.ac.kr}

\author[0009-0001-3758-9440]{Taewan Kim}
\affiliation{Astronomy Program, Department of Physics and Astronomy, Seoul National University, 1 Gwanak-ro, Gwanak-gu, Seoul 08826, Republic of Korea}

\author{Jubee Sohn}
\affiliation{Astronomy Program, Department of Physics and Astronomy, Seoul National University, 1 Gwanak-ro, Gwanak-gu, Seoul 08826, Republic of Korea}
\affiliation{SNU Astronomy Research Center, Seoul National University, 1 Gwanak-ro, Gwanak-gu, Seoul 08826, Republic of Korea}

\author{Ho Seong Hwang}
\affiliation{Astronomy Program, Department of Physics and Astronomy, Seoul National University, 1 Gwanak-ro, Gwanak-gu, Seoul 08826, Republic of Korea}
\affiliation{SNU Astronomy Research Center, Seoul National University, 1 Gwanak-ro, Gwanak-gu, Seoul 08826, Republic of Korea}
\affiliation{Australian Astronomical Optics - Macquarie University, 105 Delhi Road, North Ryde, NSW 2113, Australia}

\author{Simon C.-C. Ho}
\affiliation{Research School of Astronomy and Astrophysics, The Australian National University, Canberra, ACT 2611, Australia}
\affiliation{ARC Centre of Excellence for All-Sky Astrophysics in 3 Dimensions (ASTRO 3D)}
\affiliation{OzGrav: The Australian Research Council Centre of Excellence for Gravitational Wave Discovery, Hawthorn, VIC 3122, Australia}

\author{Denis Burgarella}
\affiliation{Aix Marseille Universit\'e, CNRS, Laboratoire d'Astrophysique de Marselle UMR 7326, 13388 Marseille, France}

\author{Tomotsugu Goto}
\affil{Institute of Astronomy, National Tsing Hua University, 101, Section 2. Kuang-Fu Road, Hsinchu, 30013, Taiwan, (R.O.C.)}

\author{Tetsuya Hashimoto}
\affil{Department of Physics, National Chung Hsing University, 
No. 145, Xingda Rd., South Dist., Taichung, 40227, Taiwan (R.O.C.)}

\author{Woong-Seob Jeong}
\affil{Korea Astronomy and Space Science Institute, 776 Daedeokdae-ro, Yuseong-gu, Daejeon 34055, Korea}

\author[0000-0001-9970-8145]{Seong Jin Kim}
\affil{Department of Physics, National Chung Hsing University, 
No. 145, Xingda Rd., South Dist., Taichung, 40227, Taiwan (R.O.C.)}

\author[0000-0001-6919-1237]{Matthew A. Malkan}
\affil{Department of Physics and Astronomy, University of California, Los Angeles, Los Angeles, CA 90095-1547, USA}

\author{Takamitsu Miyaji}
\affil{Instituto de Astronom\'ia sede Ensenada Km. 107, Carret. Tijuana-Ensenada Ensenada, 22860 BC, Mexico}

\author{Nagisa Oi}
\affil{Space Information Center, Hokkaido Information University, Nishi-Nopporo 59-2, Ebetsu, Hokkaido 069-8585, Japan}

\author{Hyunjin Shim}
\affil{Department of Earth Science Education, Kyungpook National University, 80 Daehak-ro, Buk-gu, Daegu 41566, Republic of Korea}

\author{Hyunmi Song}
\affil{Department of Astronomy and Space Science, Chungnam National University, Daejeon 34134, Republic of Korea}

\author[0000-0002-2013-1273]
{Narae Hwang}
\affil{Korea Astronomy and Space Science Institute, 776 Daedeokdae-ro, Yuseong-gu, Daejeon 34055, Korea}

\author{Byeong-Gon Park}
\affil{Korea Astronomy and Space Science Institute, 776 Daedeokdae-ro, Yuseong-gu, Daejeon 34055, Korea}

\begin{abstract}
We perform an MMT/Hectospec redshift survey of the North Ecliptic Pole Wide (NEPW) field covering 5.4 square degrees, and use it to estimate the photometric redshifts for the sources without spectroscopic redshifts. By combining 2572 newly measured redshifts from our survey with existing data from the literature, we create a large sample of 4421 galaxies with spectroscopic redshifts in the NEPW field. Using this sample, we estimate photometric redshifts of 77755 sources in the band-merged catalog of the NEPW field with a random forest model. The estimated photometric redshifts are generally consistent with the spectroscopic redshifts, with a dispersion of 0.028, an outlier fraction of 7.3 \%, and a bias of $-0.01$. We find that the standard deviation of the prediction from each decision tree in the random forest model can be used to infer the fraction of catastrophic outliers and the measurement uncertainties. We test various combinations of input observables, including colors and magnitude uncertainties, and find that the details of these various combinations do not change the prediction accuracy much. As a result, we provide a catalog of 77755 sources in the NEPW field, which includes both spectroscopic and photometric redshifts up to z$\sim$2. This dataset has significant legacy value for studies in the NEPW region, especially with upcoming space missions such as JWST, Euclid, and SPHEREx.
\end{abstract}


\section{Introduction} \label{sec1}
Estimating galaxy redshifts is one of the most important tasks in extragalactic astronomy and cosmology (e.g., \citealp{intro1, intro2, intro3}). The redshifts are generally derived from spectroscopic observations with high precisions. Large spectroscopic surveys including SDSS \citep{sdss}, HectoMAP \citep{hectomap}, and the DESI \citep{desi} provide an extensive sample of spectroscopic redshift measurements. However, obtaining a statistically large sample of spectroscopic redshifts requires a considerable amount of telescope time and resources, limiting the number of objects with available spectroscopic redshifts.

Photometric redshifts are an alternative proxy for spectroscopic redshifts (e.g., \citealp{Baum1957, Puschell1982}). The photometric redshifts of galaxies are conventionally estimated based on the comparison between the photometry (e.g., $ugriz$ across optical wavelengths) and the spectral energy distribution (SED) models (e.g., \citealp{Puschell1982, Arnouts99, Benitz00, 2000A&A...363..476B, Ilbert06}). Photometric redshifts are generally less precise with the typical uncertainty of $\delta z\sim 0.003 - 0.1$ at $z < 1$ (e.g., \citealp{Ho}), which is larger than that of spectroscopic redshifts, $\delta z \lesssim 0.0001$ at $z<1$ (e.g., \citealp{SDSS_dr16}). Photometric redshift estimation is often confused because the photometric SED can be explained by the model SED shifted with various redshifts (e.g., \citealp{Tanaka2015}). Thus, recent photometric redshift measurements are based on multi-band photometry; for example, \citet{Laigle2016} derive the photometric redshifts of the objects in the COSMOS field using 36 bands.

More recently, various machine learning techniques have been employed to refine the photometric redshift measurements \citep{photoz, Pasquet2019, Schuldt2021,Henghes2022, Luo2024}. For example, the neural network model (e.g., \citealp{2012A&A...546A..13C, 2021AJ....162..297L}) transforms photometric data into a target label with the applications of tensor (or matrix) operators and non-linear functions. It finds the optimal tensors that generate the most accurate target label with the training set \citep{1982PNAS...79.2554H, 2021AJ....162..297L}. Another example is a Gaussian process based on the assumption that an output $y_i$ for input $\boldsymbol{x}$ is described by a joint Gaussian distribution with a mean $\mu(\boldsymbol{x_i})$ and a covariance matrix $\Sigma(\boldsymbol{x})$. The Gaussian process constructs the distribution of $y$ based on the training set $(y_i, \boldsymbol{x_i})$ and predicts $y_j$ for a new input $\boldsymbol{x_j}$ \citep{2006gpml.book.....R}. As an example application of the Gaussian process to the photometric redshift estimation, \cite{2017ApJ...838....5L} combines the SED fitting with the Gaussian process on the flux-redshift space to estimate the photometric redshift.

Here, we construct a random forest model \citep{random-forest, ML} to derive photometric redshifts. The random forest model is a supervised learning that identifies an underlying function from the training set consisting of pairs of well-defined input features (e.g., photometry) and a target label (e.g., photometric redshift). The random forest technique has a simple prediction process. Our random forest photometric redshift estimation model requires a small training set and a short training time.

We test our random forest model and derive the photometric redshifts for the sources from the survey of \textit{AKARI North Ecliptic Pole Wide (NEPW) Field}. \textit{AKARI} was an infrared astronomy satellite launched in 2006 \citep{2007PASJ...59S.369M} which produced an all-sky survey. Due to its orbit, it has especially high visibility around the ecliptic pole regions. It repeatedly observed the $5.4~\text{deg}^2$ area around the NEP centered at R.A.=$18\text{h }00\text{m }00\text{s }$, Dec.=$+66^\circ\ 33'\ 38''\ $, with its nine filters from near- to mid-infrared wavelengths \citep{NEP}. This wavelength range covers the emission from polycyclic aromatic hydrocarbon (PAH) emission, a useful feature for studying dusty star-forming galaxies.

The combination of \textit{AKARI} photometric data with redshifts allows us to obtain valuable insights into the star formation history of galaxies \citep{NEP}, the physical modeling of the spectral energy distribution of galaxies \citep{kimsj19}, and the metallicity-PAH relation of star-forming galaxies \citep{2023AJ....165...31S}. Moreover, the band-merged catalog of the NEPW field \citep{NEPW_catalog} becomes an important basis for studying the properties of infrared-bright galaxies, including the infrared luminosity functions \citep{goto19}, the nuclear activity of AGN-host galaxies \citep{Chiang19, wang20, santos21, Poliszczuk21, chen21}, the star formation model of high-$z$ galaxies \citep{Barrufet20}, the infrared sources without optical counterparts \citep{toba20}, the identification of galaxy clusters \citep{huang21}, the merger-driven star formation activity \citep{kimeb21}, and optical/near-infrared identifications of X-ray sources \citep{Miyaji24}. Constructing a large sample of galaxies with redshift measurements is critical to the study of listed subjects.

Based on the multi-band photometry and the compilation of existing spectroscopy of the NEPW sources, we construct the random forest model to estimate the photometric redshifts. We also apply the random forest model to provide homogenous measurements of photometric redshifts for the sources in the NEPW field. The extended catalog of the NEPW field including the photometric redshifts will be an important basis for future studies of galaxy evolution based on optical and NIR/MIR photometry. This paper is organized as follows; we describe the photometric and spectroscopic data we use in Section \ref{sec2}. We then demonstrate our photometric redshift measurement process based on the random forest model in Section \ref{sec3}. We discuss the accuracy and significance of our photometric redshift measurements in Section \ref{sec4}. We conclude in Section \ref{sec5}.

\section{The Data Sets} \label{sec2}

\subsection{Input photometry} \label{sec2.1}

The basement for our photometric redshift measurement is multi-band photometry of the sources in the NEPW field. \textit{AKARI} observes the NEPW field with its nine filters in near- and mid-infrared ranges (e.g., \textit{N2, N3}, and \textit{N4}; \citealp{NEP, 2009PASJ...61..375L, NEPW_survey}) and detects 91681 sources. Based on the \textit{AKARI} observations, our team constructs the band-merged catalog for the NEPW field \citep{NEPW_catalog} by cross-matching the sources detected in \textit{AKARI} NEPW surveys and other follow-up observations. 

We select 26-band photometry from the NEPW band-merged catalog to measure photometric redshifts following \citet{Ho}. From the original NEPW band-merged catalog including 42-band photometry, \citet{Ho} used 26-band photometry for the photometric redshift estimates based on the spectral energy distribution (SED) fitting technique. To make a fair comparison, we also use the following 26-band photometry. In the optical range, the photometry with the highest number of observed sources comes from Subaru/HSC observations \citep{deep_hsc1, deep_hsc2}. They provide complete photometry for 77755 sources at \textit{g, r, i, z}, and \textit{Y}. In addition to Subaru/HSC photometry, there is photometry from the CFHT/MegaCam \textit{u*/g/r/i/z} \citep{cfht_obs1, cfht_obs2}, CFHT/MegaPrime \textit{u*} \citep{megaprime}, and the Maidanak/SNUCAM \textit{B/R/I} \citep{maidanak_obs}. In the infrared range, \textit{AKARI} \textit{N2, N3}, and \textit{N4} photometry is available for most sources. The infrared photometry in CFHT/WIRcam \textit{Y/J/Ks} \citep{cfht_obs2}, the KPNO/FLAMINGOS \textit{J/H} \citep{2014ApJS..214...20J}, \textit{Spitzer} IRAC1/2 \citep{2004AAS...204.3301W, wise, spitzer}, and WISE W1/W2 \citep{2010AJ....140.1868W, wise} bands are also available. We note that most sources in the NEPW band-merged catalog miss photometry in some bands because the depths and spatial coverages of the observations are different. Table 1 (and Figure 3) in \citet{Ho} summarizes the photometric bands we use for our photometric redshift estimation. 

\subsection{Spectroscopic data and redshift measurement}\label{sec2.2}

We compile spectroscopic redshifts from the previous observations \citep{specz1, specz2, specz3, specz4, specz5, specz6, specz7}. We also add new spectroscopic redshifts based on our own spectroscopic observations for the NEPW field. 

\begin{figure}[t]
    \includegraphics[width=.48\textwidth]{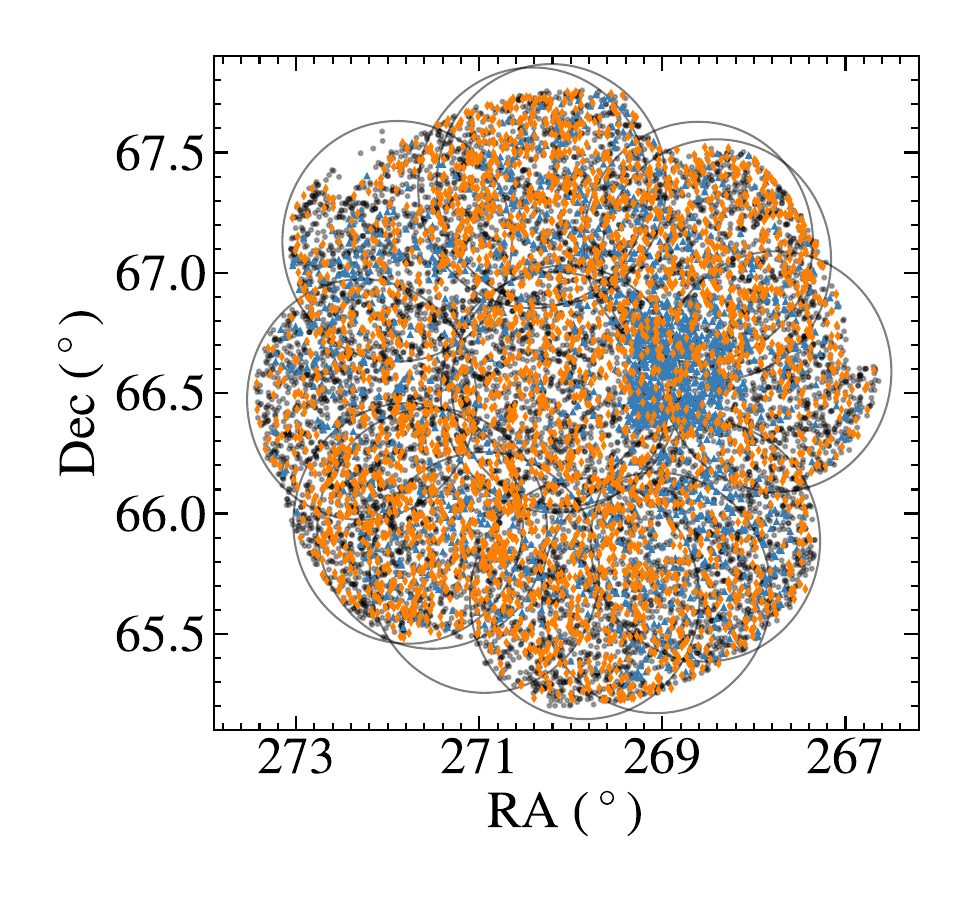}
    \caption{The field of view of our own spectroscopic survey of the NEPW field. The gray dots indicate the sources identified in the infrared range in the AKARI NEPW field survey. We plot 10 percent of all samples detected by \textit{AKARAI} for clarity. Blue dots show the sources with spectroscopic redshifts measured in previous studies. Orange dots display the sources with spectroscopic redshifts measured from our MMT/Hectospec survey. The large black circles indicate the MMT/Hectospec survey fields.} 
    \label{fig1}
\end{figure}

\begin{deluxetable*}{ccccccc}
\tablecaption{MMT/Hectospec Observations on NEPW field} \label{tab1}
\tablehead{\colhead{Field} & \colhead{R.A.} & \colhead{Decl.} & \colhead{Date} & \colhead{Exposure} &\colhead{Number of} &\colhead{Number of} \\ 
\colhead{} & \colhead{(deg, J2000)} & \colhead{(deg, J2000)} & \colhead{} & \colhead{(seconds)} & \colhead{Targets} &\colhead{Measured Redshifts}}
\startdata
NEPc20\_2 & 269.068 & 65.67054 & 2020 Oct 7 & 3600 & 260 & 160 \\
NEPc20\_1 & 270.168 & 66.50360 & 2020 Oct 11 & 3600 & 258 & 221 \\
NEPc21\_1 & 271.897 & 67.13100 & 2021 Oct 2 & 3352 & 248 & 148 \\
NEPc21\_3 & 270.943 & 65.75491 & 2021 Oct 2 & 3600 & 251 & 153 \\
NEPc21\_2 & 268.411 & 67.05496 & 2021 Oct 3 & 3600 & 259 & 214 \\
NEPc21\_4 & 271.514 & 65.93796 & 2021 Oct 3 & 2880 & 258 & 130 \\
NEPc21\_5 & 270.212 & 67.36748 & 2021 Nov 12 & 3600 & 254 & 194 \\
NEPa22\_1 & 267.752 & 66.58976 & 2022 Mar 2 & 4140 & 259 & 168 \\
NEPa22\_2 & 272.282 & 66.47379 & 2022 Apr 4 & 3600 & 260 & 210 \\
NEPa22\_3 & 269.848 & 65.64594 & 2022 Apr 25 & 3600 & 256 & 102 \\
NEPa22\_4 & 268.531 & 65.88472 & 2022 Apr 28 & 4800 & 261 & 205 \\
NEPa22\_5 & 270.413 & 67.35335 & 2022 Apr 29 & 4800 & 255 & 199 \\
NEPa22\_6 & 271.777 & 65.95914 & 2022 May 3 & 4860 & 256 & 183 \\
NEPa22\_7 & 268.609 & 67.12775 & 2022 May 4 & 3960 & 225 & 196 \\
NEPa22\_8 & 270.048 & 66.53124 & 2022 May 4 & 2400 & 254 & 132 \\
\enddata
\end{deluxetable*}

We first obtain 1985 spectroscopic redshifts from the NEPW band-merged catalog \citep{NEPW_catalog}. Most of the spectroscopic redshifts in the catalog are originated from \cite{specz1}. \cite{specz1} observe the NEPW field with MMT/Hectospec and WIYN/Hydra. Based on the visual inspection, \citet{specz1} provide 1796 reliable redshift measurements, including 6 Milky Way stars, 233 AGNs, 1522 galaxies, and 35 unidentified sources due to the low signal-to-noise ratios. The other spectroscopic redshifts are from Keck/DEIMOS \citep{specz2, specz3, specz4}, Gran Telescope Canarias (GTC)/OSIRIS \citep{specz5}, Subaru/FMOS \citep{specz6}, and AKARI/SPICY \citep{specz7} observations.

We conduct our own spectroscopic survey with MMT/Hectospec to increase the number of spectroscopic redshifts in the NEPW field. Our survey was carried out from 2020 to 2022 as a part of the K-GMT science program (PI: Ho Seong Hwang). Hectospec is a multi-object optical spectrograph with 300 fed fibers for MMT \citep{hectospec}. It covers the wavelength range of 3500 \AA\ - 10000 \AA\, with a spectral resolution of R$\sim$1000-2000. The main targets of our survey were \textit{AKARI} $9\ \mu\text{m}$ selected galaxies with $r$-band magnitudes brighter than 21 mag. We additionally obtained spectra for the galaxies that are bright in \textit{r}-band, but not detected in $9\ \mu\text{m}$. We reduced the Hectospec spectra based on the HSRED reduction pipeline\footnote{The description on the pipeline at \url{http://www.mmto.org/hsred-reduction-pipeline/}}. 

Table \ref{tab1} summarizes our spectroscopic survey of the NEPW field with MMT/Hectospec. In our 15 visits to the NEPW, we observed 3835 sources ($\sim$ 250 sources per field). Figure \ref{fig1} illustrates the spatial distribution of the NEPW sources. Our survey uniformly covers the NEPW regions and almost triples the number of sources with optical spectra.

To measure the redshift from the reduced spectrum, we develop a new \texttt{Python} package, \texttt{RVSNUpy}, based on a cross-correlation technique (T. Kim et al. 2025 in preparation). The code implements the algorithm that was originally used for \texttt{IRAF/RVSAO} \citep{rvsao}; it first subtracts the continuum of a given spectrum and cross-correlates the continuum-subtracted spectrum against the various template spectra, including absorption- and emission-line dominated templates in the rest frame. Because the cross-correlation signal is maximized when the phases of the template and the given spectrum are most similar, \texttt{RVSNUpy} determines the peak of the cross-correlation signal as the redshift of the given spectrum. \texttt{RVSNUpy} also provides the $r$-value from \citet{td79}, which indicates the significance of the cross-correlation signals. We explain the details of \texttt{RVSNUpy} in our forthcoming paper (T. Kim et al. in preparation).

We measure the redshifts of the observed sources with \texttt{RVSNUpy} based on all prepared templates and determine the final redshifts as follows. First, we select reliable redshift measurements with $r \gtrsim 4$ derived based on absorption-line-dominated templates and $r \gtrsim 20$ derived based on emission-line-dominated templates. Second, we prioritize the measurements based on the absorption-line-dominated templates over the emission-line-dominated templates because the absorption lines originate from the stars in the galaxies. Lastly, we prioritize the measurements with the higher $r$-value. We finally obtained 2572 reliable redshift measurements. 

\begin{figure*}[t]
    \centering
    \includegraphics[width=\textwidth]{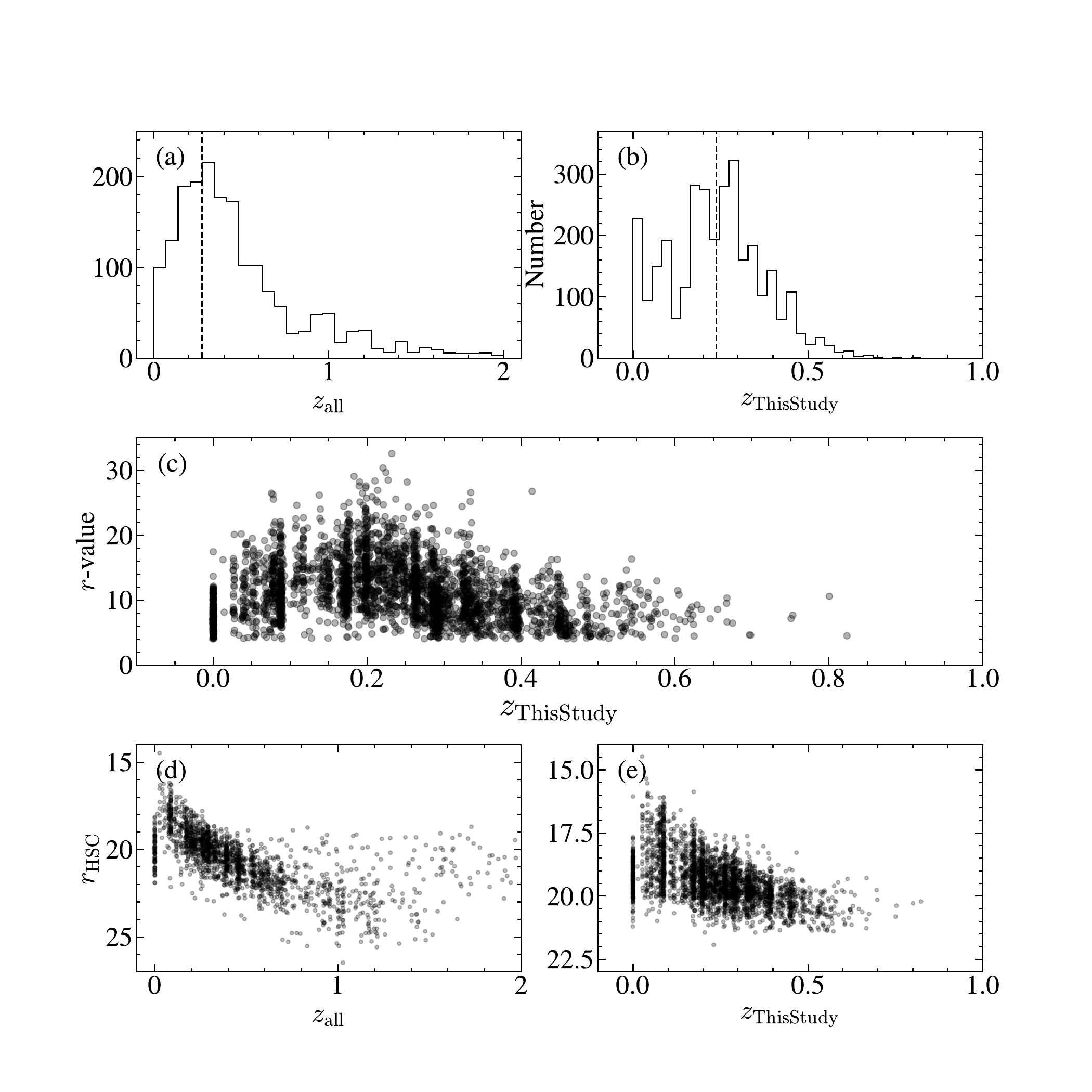}
    \caption{Properties of spectroscopically observed sources in our NEPW sample as a function of spectroscopic redshifts. The top panel shows the distribution of (a) all available spectroscopic redshifts and (b) spectroscopic redshifts measured in this study. The vertical dashed line indicates the median of the redshifts. The middle panel shows the \texttt{RVSNUpy} $r$-value distribution as a function of the spectroscopic redshift measured in this study. The bottom panel is for the Subaru/HSC r-band magnitudes as a function of (d) all available spectroscopic redshifts and (e) the spectroscopic redshifts measured in this study. }
    \label{fig2}
\end{figure*}

Figure \ref{fig2} (a) shows the distributions of 4467 spectroscopic redshifts in the combined NEPW spectroscopic catalog. Figure \ref{fig2} (b) displays the distribution of 2572 redshifts we obtained from our own spectroscopic survey. The new redshifts are mostly at $z < 1$. Figure \ref{fig2} (c) shows the $r-$value distribution as a function of redshift. Our new redshift measurements are all reliable with $r > 4$. Figure \ref{fig2} (d) and (e) show the Subaru/HSC $r-$band magnitude as a function of spectroscopic redshift. The vertical concentrations of galaxies at similar redshifts indicate clustering of galaxies. Our spectroscopic sample includes galaxies in $15 < r ~({\rm mag}) < 26$. The median redshift of our spectroscopic sample is 0.274. For our photometric redshift estimation, we select 4421 galaxies at $z < 2$ because the number of high-z galaxies is too small for our random forest applications. 

\section{Photometric redshift with a random forest} \label{sec3}

We build a random forest machine learning model to estimate photometric redshifts. We first describe the decision tree which is a basis for our random forest model in Section \ref{sec3.1}. In Section \ref{sec3.2}, we construct the random forest model to estimate the photometric redshifts based on our NEPW photometric dataset. In Section \ref{sec3.3}, we compare the photometric redshift measurements based on our random forest model with the redshift measurements based on spectroscopy and SED-fitting. In Section \ref{sec3.4}, we describe methods to determine the reliabilities and uncertainties of the photometric redshift measurements. In Section \ref{sec3.5}, we present our value-added NEPW photometric redshift catalog based on our random forest model.

\subsection{Decision tree}\label{sec3.1}

The decision tree \citep{ML} is a supervised learning that makes a prediction based on the partition of an input data space. The decision tree is often employed for both classification and regression. Here, we briefly explain the regression process based on the decision tree used for estimating photometric redshifts.

Developing the decision tree requires a training set containing input features and a target value. In our photometric redshift estimation, the input features are multi-band photometry of the galaxies in the NEPW field, and the target values are their spectroscopic redshifts. The decision tree first divides the training set into two groups that have similar target values. Specifically, the decision tree computes the similarity of the target values in the two groups, which we refer to as RSS: 
\begin{equation}
    \text{RSS}(J) = \sum^J_{j=1}\sum_{i\in S_j}(z_i-\overline{z}_j)^2.
    \label{RSS}
\end{equation}
Here, $z_{i}$ is the target value of $i-$th element in a group divided by the decision tree and $\overline{z}_{j}$ is the mean of $z_{i}$ in the group. For the first step of the decision tree ($J=1$), there is only one group (i.e., $S_{j=1}$) for the training set. As we increase the depth ($J$) of the tree, there are more and smaller groups, like the branches stemming out from a single tree.

At each depth, the decision tree determines the boundary in the input feature space for grouping that minimizes the RSS. In other words, the decision tree tests different groupings by changing the boundary in the input feature space and finds the grouping with the lowest RSS value. In our case, the decision tree tests grouping based on magnitude, magnitude uncertainty, and color distributions in multi-band photometry, selecting the grouping that minimizes the RSS value. By repeating this grouping, the decision tree splits the training sample into more groups that have similar target values within a narrower range.

As the depth of the decision tree increases, the groups capture more detailed relations between the input features and the target values, enabling more accurate prediction. However, the decision tree grouping becomes too specific to the training data when the depth is too large (i.e., overfitting). Then, the model prediction based on the decision tree becomes less accurate for the new data with slightly different input feature space. Thus, the optimal depth of the decision tree needs to be determined empirically depending on the purpose of the model (see Section \ref{sec3.2}).

Once the decision tree is constructed, the prediction for the target values is available with a new input dataset with the same input features. In our case, the new input dataset including the photometry of galaxies goes through the decision tree and returns the spectroscopic redshift prediction based on the decision tree.

\subsection{Construction of our random forest model} \label{sec3.2}

\begin{figure*}[t]
    \centering
    \includegraphics[width=\textwidth]{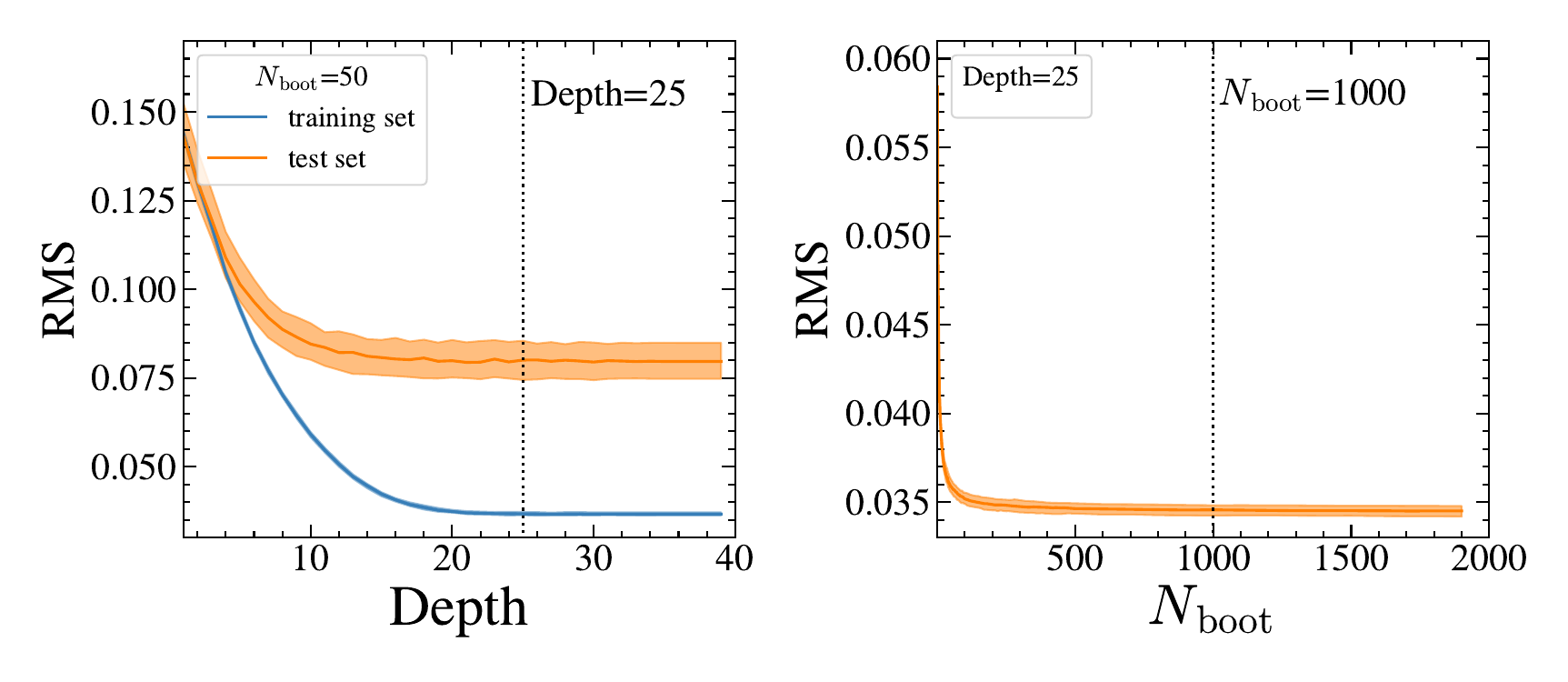}
    \caption{(Left) RMS of the redshift measurements for training (blue) and test (orange) sets as a function of the depth. The shaded region displays the $1\sigma$ distributions of RMS. The dashed vertical line denotes the depth of 25. (Right) RMS of the test set as a function of the number of bootstrapped sub-training sets. The dashed vertical line denotes 1000 of the number of bootstrapped sub-training sets.}
    \label{fig3}
\end{figure*}

Random forest is a machine learning algorithm that combines multiple decision trees to make a robust prediction. A single decision tree generally has an accuracy lower than other machine learning models because of its simple structure. Combining multiple models to construct a structured model can help improve prediction accuracy. The random forest bootstraps the training set to construct sub-training sets and generates a decision tree for each sub-training set \citep{random-forest, ML}. Each decision tree uses only randomly selected features at each partitioning step. The random selection of features `decorrelates' the decision trees and improves regression performance. The random forest then returns the average of outputs from all decision trees as its final output.

We construct a random forest model to predict photometric redshifts. We use the \texttt{scikit-learn} \citep{scikit-learn} and follow the approach\footnote{https://www.astroml.org/book\_figures/chapter9/fig\_photoz\_forest.html\label{mlexample}} from \cite{astroML}. The input photometric dataset lists 56 input features, including the 26 magnitudes and their uncertainties, and four Subaru/HSC colors (i.e., {\textit{g-r, r-i, i-z, z-Y}}). We use colors only from Subaru/HSC because they provide complete photometry for most sources.

Many galaxies in the NEPW band-merged catalog \citep{NEPW_catalog} do not have full 26 magnitudes and their uncertainties. We replace the missing photometry data with a dummy value of 1000 to maintain the dimensionality of the input data \citep{dummyvalue}. The prediction with missing values replaced by a dummy value is straightforward for a random forest model compared to other algorithms (e.g., neural networks). We choose an extreme dummy value that significantly differs from the actual photometric measurements in the input feature space. The random forest model recognizes these dummy values as less informative and makes predictions without relying on them.

Construction of our random forest model requires the determination of three hyperparameters: the number of features used for partitioning, the depth of the decision trees, and the number of bootstrapped sub-training sets. For the number of features used for partitioning, we use the square root of the number of all features because it generally yields the best performance in the regression based on a random forest \citep{ML}. For our random forest model, the number of features used for partitioning is $7\approx\sqrt{56}$.

We determine the depth of the decision trees and the number of bootstrapped sub-training sets based on the performance of the photometric redshift estimation for 4421 NEPW sources with spectroscopic redshifts at $z < 2$. We randomly select 10 percent (i.e., 442 sources) of the sources with spectroscopic redshifts as a test set and 90 percent of them as a training set. We predict the output photometric redshifts for the sources in the test set based on our random forest model trained with the training set. We compute the root mean square (RMS) errors between the predicted and spectroscopically measured redshifts. We repeat this test 100 times.

Figure \ref{fig3} shows the RMS of the predicted (i.e., photometric) redshift measurements as a function of the hyperparameters. The left panel of Figure \ref{fig3} displays the RMS of the training and test sets as a function of the depth of the decision trees. We fix the number of bootstrapped sub-training sets as 1000 here. As the depth of the tree increases, the RMS of the training and test set decreases. The RMS converges after the depth of 25 for both the training and test set. Thus, we choose 25 as the depth of the decision trees for our random forest model.

The right panel of Figure \ref{fig3} shows the RMS of the test set as a function of the number of bootstrapped sub-training sets ($N_{\rm boot}$). For this test, we fix the depth of the decision trees as 25. The RMS decreases rapidly as the number of bootstrapped sub-training sets increases and converges after $N_{\rm boot}=1000$. Thus, we choose 1000 as the number of bootstrapped sub-training sets for our random forest model.

\subsection{Comparison with spectroscopic redshifts and photometric redshifts derived from SED fitting}
\label{sec3.3}

\begin{figure*}[t]
    \centering
    \includegraphics[width=\textwidth]{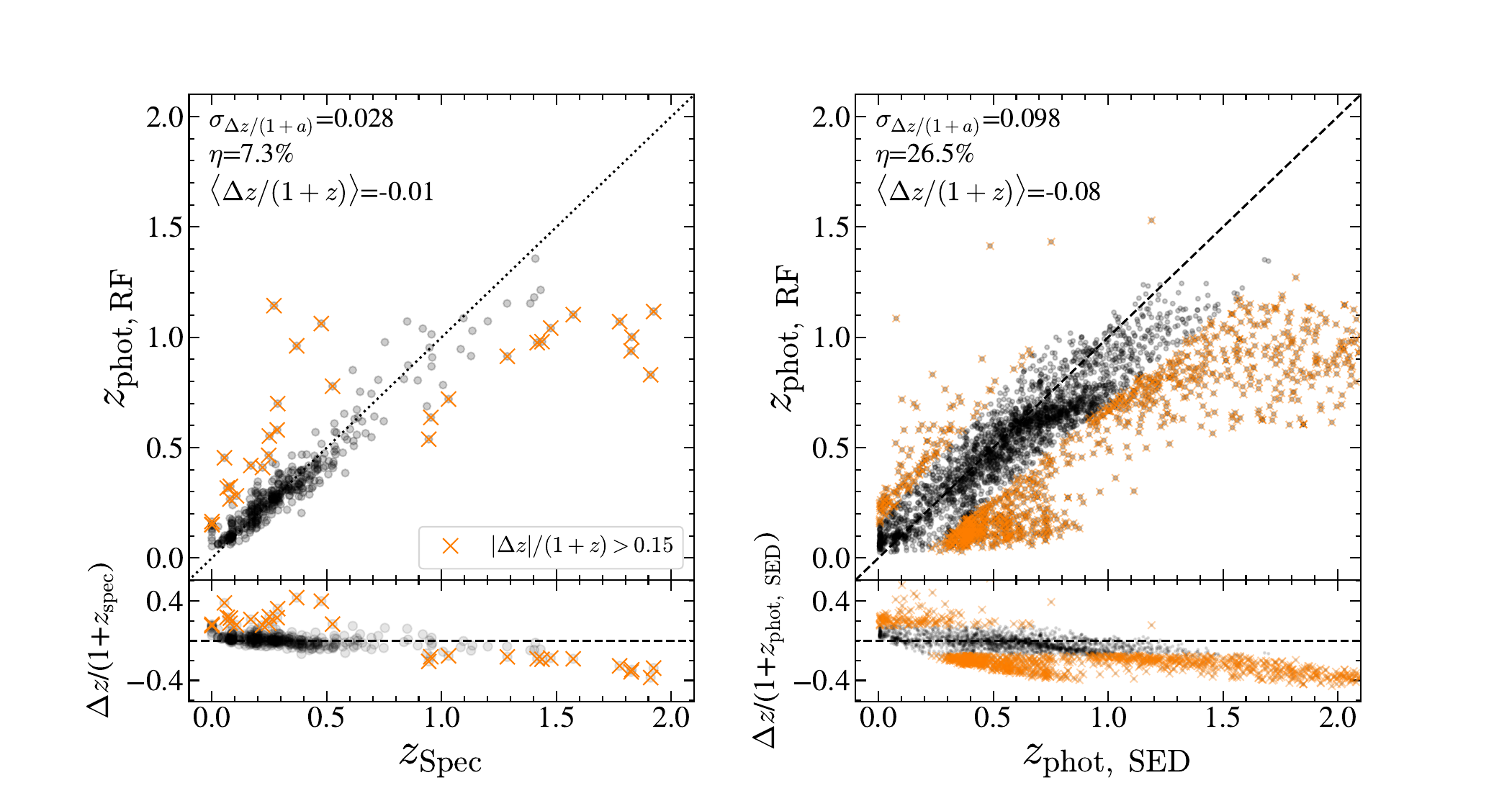}
    \caption{(Left) Comparison between the photometric redshifts measured with our random forest model ($z_{\rm phot, RF}$) and the spectroscopic redshifts ($z_{spec}$). Orange cross symbols indicate the catastrophic outliers with $|\Delta z|/(1+z)>0.15$. The quoted numbers are the accuracy metrics of the $z_{\rm phot, RF}$. (Right) Photometric redshifts measured from the SED fit ($z_{\rm phot, SED}$) vs. $z_{\rm phot, RF}$. For clarity, we display only 5 percent of the data.}
    \label{fig4}
\end{figure*} 

We evaluate the performance of our random forest photometric redshifts ($z_{\rm phot, RF}$) based on the comparison with spectroscopic redshifts ($z_{\rm spec}$). For this test, we use 4421 sources with spectroscopic redshifts for training and testing our random forest model. We randomly select 442 sources ($\sim 10\%$) with $z_{\rm spec}$ as a test set (Section \ref{sec3.2}). We train our random forest model with the remaining. We repeat the test 100 times for different test sets. 

We quantify the accuracy of $z_{\rm phot, RF}$ of the test set with three different accuracy metrics. The first metric is the dispersion of the redshift difference ($\sigma_{\Delta z/(1+z)}$). Here, the dispersion of the redshift difference corresponds to $1.48\times\text{MAD}(|\Delta z|)$, where $\Delta z=z_{\rm phot,RF}-z_{\rm spec}$ and MAD is the median absolute deviation (e.g., \citealp{Ho}). The second metric is the fraction of catastrophic outliers ($\eta$); the sources with $|\Delta z/(1+z_{\rm spec})|>0.15$ are classified as outliers. The third metric is the bias ($\langle\Delta z/(1+z)\rangle$), indicating the mean of the normalized difference between photometric and spectroscopic redshifts. 

The left panel of Figure \ref{fig4} compares $z_{\rm spec}$ and $z_{\rm phot, RF}$ derived from one example test set. We quote the median of the three accuracy metrics in Figure \ref{fig4}. The photometric redshifts generally agree with the spectroscopic redshifts with a very small offset (bias $\sim -0.01$), except for 30 outliers (orange crosses, $\sim7.3$ \%). Our random forest model estimates photometric redshifts with smaller redshift difference dispersion and lower catastrophic outlier rates than those from SED fitting ($\sigma_{\Delta z/(1+z)}=0.049$ and $\eta$=8.9\%, \citealp{Ho}).   

We also compare the $z_{\rm phot, RF}$ and the photometric redshift measurements based on the SED fitting ($z_{\rm phot, SED}$, \citealp{Ho}). For this test, we train our random forest model using all sources with spectroscopic redshifts. The right panel of Figure \ref{fig4} compares $z_{\rm phot, RF}$ and $z_{\rm phot, SED}$. The accuracy metrics are also computed based on the comparison between  $z_{\rm phot, RF}$ and $z_{\rm phot, SED}$. The photometric redshifts based on two different techniques are generally consistent with each other at $z < 0.8$. The accuracy metrics indicate that $\sim75$ percent of $z\mathrm{_{phot, RF}}$ is consistent with $z\mathrm{_{phot, SED}}$ with a typical dispersion of $\sigma_{\Delta z/(1+z)}\sim0.1$ at $z < 0.8$. $z_{\rm phot, RF}$ tend to be smaller than $z\mathrm{_{phot, SED}}$ at $z\mathrm{_{phot, SED}} > 0.8$. We explore the reliability of $z_{\rm phot, RF}$ and $z_{\rm phot, SED}$ at $z > 0.8$ in Section \ref{sec3.4}. 

\subsection{Reliability and uncertainty of photometric redshift estimation}
\label{sec3.4}

The random forest returns only a single prediction value unlike many algorithms providing measurement uncertainties in their predictions (e.g., \citealp{D'Isanto2018, Eriksen2019, Soo2021}). To estimate the reliability and uncertainty of the $z_{\rm phot, RF}$, we use the variance of the predictions from the decision trees in our random forest model. Because the random forest model makes predictions based on the combination of multiple decision trees, we expect the variance of the predictions from the decision trees to be related to the confidence of the photometric redshift measurement.

\begin{figure}[t]
    \centering
    \includegraphics[width=.5\textwidth]{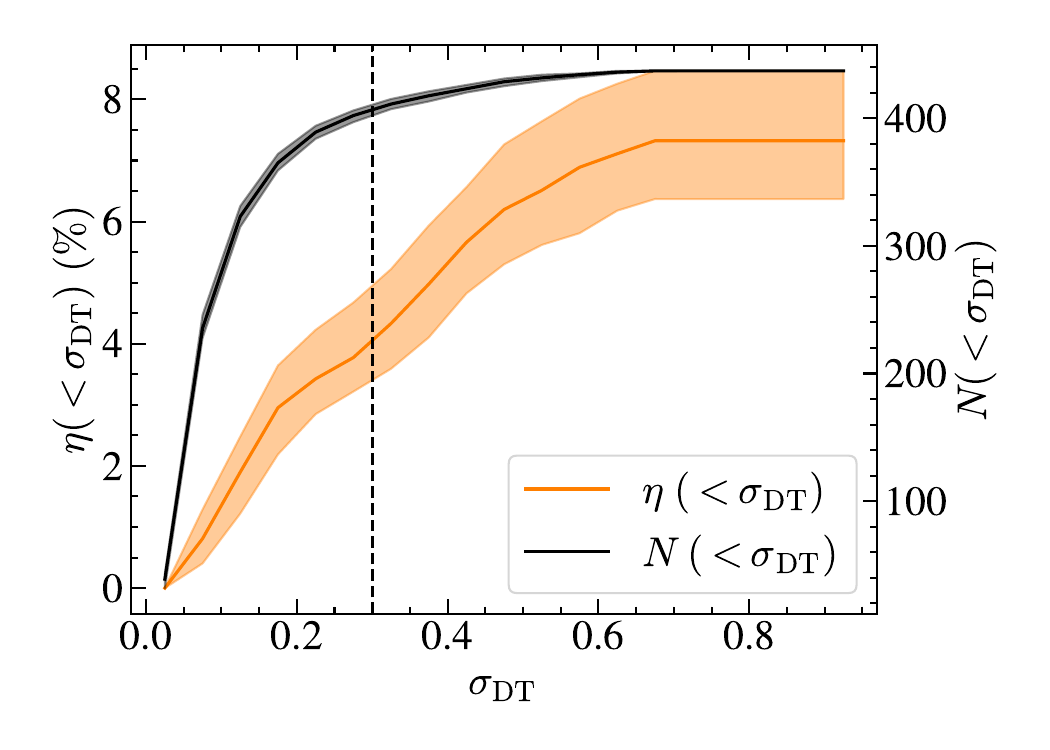}
    \caption{Cumulative fraction of catastrophic outliers and cumulative number of photometric measurements as a function of $\sigma_{\rm DT}$. The orange line and shaded region indicate the cumulative fractions of catastrophic outliers and their $1\sigma$ scatter. The black line and shaded region indicate the cumulative numbers of photometric redshift measurements and their $1\sigma$ scatter. The vertical dashed lines indicate the boundary we determine the reliability of $z_{\rm phot, RF}$.}
    \label{fig5}
\end{figure}

We analyze a cumulative fraction of catastrophic outliers (i.e., $|\Delta z|/(1+z)>0.15$) from 100 randomly selected test sets (hereafter the uncertainty test sample) as a function of the standard deviation of the predictions from decision trees ($\sigma_{\rm DT}$). Our random forest model has 1000 decision trees. For a single test set, each decision tree yields its own prediction for the photometric redshift. Thus, $\sigma_{\rm DT}$ indicates the standard deviation of 1000 predictions for the photometric redshift.  

Figure \ref{fig5} shows the cumulative fraction of the catastrophic outliers and the cumulative number of photometric measurements as a function of $\sigma_{\rm DT}$. The outlier fraction increases with $\sigma_{\rm DT}$. For $\sigma_{\rm DT} < 0.3$, 98 percent of the uncertainty test set is included and the fraction of catastrophic outliers remains below five percent. We thus conclude $\sigma_{\rm DT}=0.3$ as a criterion to determine the reliability of our photometric redshift measurement.

\begin{figure}[t]
    \centering
    \includegraphics[width=.5\textwidth]{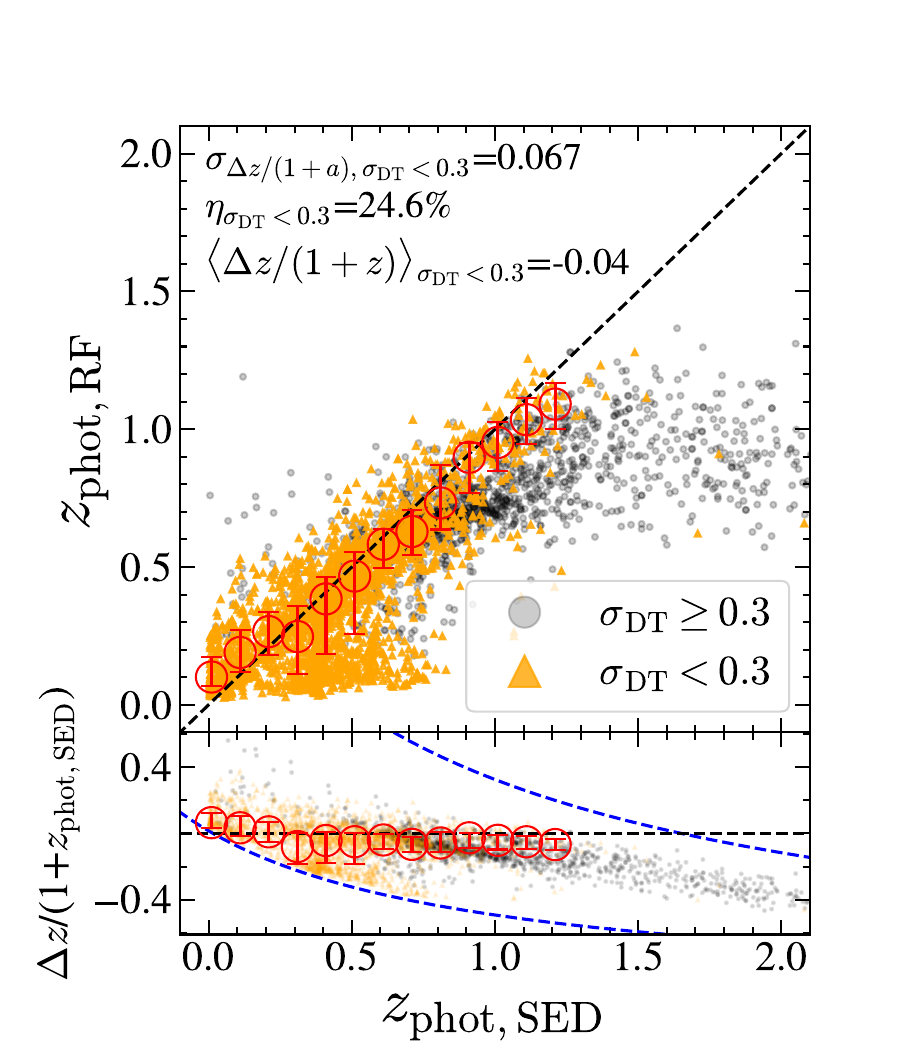}
    \caption{Comparison between  $z_{\rm phot, RF}$ and $z_{\rm phot, SED}$ for two groups with $\sigma\mathrm{_{DT}} < 0.3$ (orange triangles) and with $\sigma\mathrm{_{DT}} > 0.3$ (gray circles). For clarity, we display only 5 percent of the data. Red circles mark the median $z_{\rm phot, RF}$ with $\sigma\mathrm{_{DT}} < 0.3$ at each $z_{\rm phot, SED}$ bin. In the lower panel, blue dashed lines indicate the maximum and minimum offset between $z_{\rm phot, RF}$ and $z_{\rm phot, SED}$ at each $z_{\rm phot, SED}$ introduced because of the maximum spectroscopic redshift ($z_{\rm spec} <2$) of our training sample.}
    \label{fig6}
\end{figure}

Figure \ref{fig6} shows the comparison between $z_{\rm phot, RF}$ and $z_{\rm phot, SED}$ for the two groups separated at $\sigma\mathrm{_{DT}} = 0.3$. In general, $z_{\rm phot, RF}$ with $\sigma\mathrm{_{DT}}<0.3$ are consistent with $z_{\rm phot, SED}$. The objects with inconsistent $z_{\rm phot, RF}$ and $z_{\rm phot, SED}$ have $\sigma\mathrm{_{DT}}$ larger than 0.3. At $z_{\rm phot, SED} < 0.8$, there is a group of objects where $z_{\rm phot, RF}$ are significantly lower than  $z_{\rm phot, SED}$ ($\Delta z/(1+z) < -0.1$). For these objects, $z_{\rm phot, RF}$ are consistent with $z_{\rm spec}$ unlike $z_{\rm phot, SED}$, indicating that $z_{\rm phot, RF}$ more reliable. 

\begin{figure*}[t]
    \centering
    \includegraphics[width=1.\textwidth]{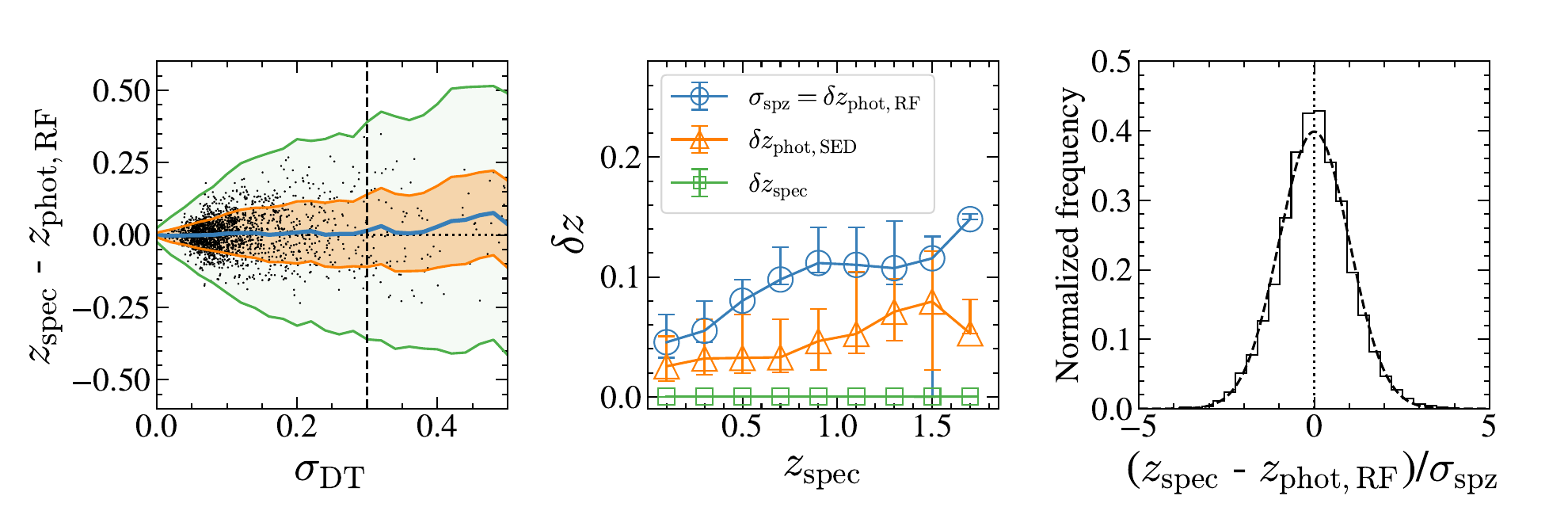}
    \caption{(Left) Difference between $z_{\rm spec}$ and $z_{\rm phot, RF}$ as a function of $\sigma_{\rm DT}$. For clarity, we display only 5 percent of the data points. The blue line indicates the mean distribution, and the orange and green area corresponds to $1\sigma_{\text{spz}}$ and $3\sigma_{\text{spz}}$ distributions. The vertical dashed lines indicate the $\sigma_{\rm DT}=0.3$. (Middle) The distributions of $\sigma_{\rm spz}$ (blue circle), the uncertainties of $z_{\rm phot, SED}$ (orange triangle), and the uncertainties of $z_{\rm spec}$ (green rectangular) across $z_{\rm spec}$ bins. The error bars indicate $1\sigma$ scatter. (Right) Normalized differences between spectroscopic and photometric redshifts using $\sigma_{\text{spz}}$. The dashed curve indicates the Gaussian curve fitted to the normalized differences.}
    \label{fig7}
\end{figure*}

We also devise a method to derive the uncertainties of $z_{\rm phot, RF}$ using $\sigma\mathrm{_{DT}}$. We examine the difference between $z_{\rm spec}$ and $z_{\rm phot, RF}$ as a function of $\sigma\mathrm{_{DT}}$ for the uncertainty test sample (the left panel of Figure \ref{fig7}). The blue solid line displays the mean redshift offset. The mean difference remains close to zero as long as there is enough data in the bin (i.e., $\sigma_{\rm DT} < 0.3$), indicating no systematic between spectroscopic and photometric redshifts. The orange and green shaded regions mark the $1\sigma$ and $3\sigma$ boundaries. This trend indicates that the range of the $z_{\rm spec}$ and $z_{\rm phot, RF}$ offset increases as a function of $\sigma\mathrm{_{DT}}$. We define the $1\sigma$ of the $(z_{\rm spec} - z_{\rm phot, RF})$ as a function of $\sigma_{\rm DT}$ as $\sigma_{\rm spz} (\sigma_{\rm DT})$, corresponding to $z_{\rm phot, RF}$ uncertainty at given $\sigma_{\rm DT}$. The middle panel of Figure \ref{fig7} displays the distributions of $\sigma_{\rm spz}$, the uncertainties of $z_{\rm phot, SED}$ ($\delta z_{\rm phot, SED}$), and the uncertainties of $z_{\rm spec}$ ($\delta z_{\rm spec}$) as a function of $z_{\rm spec}$. The $\sigma_{\rm spz}$ tend to be larger than $\delta z_{\rm phot, SED}$ for given $z_{\rm spec}$ and increase as $z_{\rm spec}$ increase. We note that the definitions of $\sigma_{\rm spz}$ and $\delta z_{\rm phot, SED}$ are different, thus comparing them is not straightforward. We discuss the performance of $z_{\rm phot, RF}$ in comparison with that of $z_{\rm phot, SED}$ in Section \ref{sec4.3}.

To justify the use of $\sigma\mathrm{_{spz}}$ as the measurement uncertainty, we analyze the distribution of the $(z_{\rm spec} - z_{\rm phot, RF})$ normalized by the $\sigma_{\text{spz}}$ for the uncertainty test sample. The distribution of measurements divided by their uncertainties follows the normal distribution: $(z\mathrm{_{phot, RF}}-z\mathrm{_{spec}})/\sqrt{\delta z\mathrm{_{phot, RF}}^2+\delta z\mathrm{_{spec}}^2}\sim\mathcal{N}(0,1)$. Here, $\delta z\mathrm{_{phot, RF}}$ indicates the uncertainty of $z\mathrm{_{phot, RF}}$. Because the uncertainties of the spectroscopic redshifts are much smaller than those of the photometric redshifts, we expect $(z\mathrm{_{phot, RF}}-z\mathrm{_{spec}})/\sigma\mathrm{_{spz}}\sim\mathcal{N}(0,1)$ if $\sigma\mathrm{_{spz}}\approx\delta z\mathrm{_{phot, RF}}$. The right panel of Figure \ref{fig7} shows the distribution of the normalized differences between the spectroscopic and photometric redshifts. The distribution follows a normal distribution. Therefore, we include $\sigma_{\text{spz}}$ for each source as the measurement uncertainty for the photometric redshift estimates.

\subsection{A photometric redshift catalog of the NEPW field}
\label{sec3.5}
\begin{deluxetable}{ccr}
    \label{tab2}
    \tablecaption{Summary of our value-added NEPW redshift catalog.}
    \tablehead{Parameter & \phantom{abcdefghijk} & Value}
    \startdata
    Survey Area \tablenotemark{a} (deg$^2$) & &5.6 \\
    N$_\text{source}$\tablenotemark{b} & &77755 \\
    N$_{\sigma\mathrm{_{DT}}<0.3}$\tablenotemark{c}  & & 52167 \\
    N$_\text{spec-z, all}$\tablenotemark{d}  & &4467 \\
    \enddata
    \tablenotetext{a}{The survey area}
    \tablenotetext{b}{The number of sources included in the catalog}
    \tablenotetext{c}{The number of sources with $\sigma\mathrm{_{DT}}<0.3$}
    \tablenotetext{d}{The number of all spectroscopic redshifts available in the catalog}
\end{deluxetable}

\begin{deluxetable*}{ccccccccccc}
    \label{tab3}
    \tablecaption{Redshifts in the NEPW field.}
    \tablehead{\colhead{NEPW\_ID} \tablenotemark{a} & \colhead{R.A.} & \colhead{Decl.} & \colhead{r$_\text{HSC}$} & \colhead{specz} & \colhead{specz\_err}\tablenotemark{b} & \colhead{specz source} \tablenotemark{c} & \colhead{$r$-value}\tablenotemark{b} & \colhead{photz\_RF} & \colhead{photz\_err} \tablenotemark{d} & \colhead{$\sigma\mathrm{_{DT}}$} \\
    \colhead{} & \colhead{(deg)} & \colhead{(deg)} & \colhead{(mag)}}

    \startdata
    119624 & 268.6480 & 67.2392 & 23.328 & -9.999 & -9.999 & 9 & -9.99 & 0.6653 & 0.1412 & 0.41 \\
    119626 & 268.5046 & 67.2395 & 20.937 & -9.999 & -9.999 & 9 & -9.99 & 0.4170 & 0.0802 & 0.14 \\
    119627 & 268.3835 & 67.2417 & 22.266 & -9.999 & -9.999 & 9 & -9.99 & 0.5352 & 0.1039 & 0.23 \\
    119630 & 268.3894 & 67.2466 & 19.283 & -9.999 & -9.999 & 9 & -9.99 & 0.0501 & 0.0684 & 0.10 \\
    119631 & 268.3978 & 67.2442 & 24.597 & -9.999 & -9.999 & 9 & -9.99 & 0.9397 & 0.1524 & 0.43 \\
    119633 & 268.7164 & 67.2457 & 23.272 & -9.999 & -9.999 & 9 & -9.99 & 0.6902 & 0.1343 & 0.40 \\
    119634 & 268.7115 & 67.2433 & 23.319 & -9.999 & -9.999 & 9 & -9.99 & 0.6557 & 0.1343 & 0.39 \\
    119635 & 268.4547 & 67.2798 & 16.688 & 0.0900 & -9.999 & 1 & -9.99 & 0.0858 & 0.0081 & 0.01 \\
    119637 & 268.4605 & 67.2593 & 18.261 & -9.999 & -9.999 & 9 & -9.99 & 0.0640 & 0.0327 & 0.05 \\
    119638 & 268.4791 & 67.2695 & 20.168 & -9.999 & -9.999 & 9 & -9.99 & 0.1038 & 0.0941 & 0.17 \\
    119639 & 268.4817 & 67.2489 & 20.722 & -9.999 & -9.999 & 9 & -9.99 & 0.1386 & 0.0941 & 0.18 \\
    119640 & 268.4627 & 67.2655 & 19.833 & 0.3234 & 0.0001 & 2 & 9.40 & 0.3057 & 0.0327 & 0.06 \\
    119641 & 268.4778 & 67.2674 & 21.167 & -9.999 & -9.999 & 9 & -9.99 & 0.4055 & 0.0941 & 0.18 \\
    119642 & 268.4684 & 67.2722 & 20.870 & -9.999 & -9.999 & 9 & -9.99 & 0.3616 & 0.0979 & 0.19 \\
    119643 & 268.4848 & 67.2570 & 21.402 & -9.999 & -9.999 & 9 & -9.99 & 0.5406 & 0.1156 & 0.26 \\
    119646 & 268.4622 & 67.2636 & 21.826 & -9.999 & -9.999 & 9 & -9.99 & 0.5284 & 0.0941 & 0.17 \\
    \enddata
    \tablenotetext{a}{ID listed in the AKARI NEPW catalog of \cite{NEPW_catalog}.}
    \tablenotetext{b}{The uncertainties of spectroscopic redshifts and the $r$-values from \texttt{RVSNUpy}. We list dummy values (-9.999 or -9.99) for those objects without \texttt{RVSNUpy} redshifts.}
    \tablenotetext{c}{Sources of spectroscopic redshifts: 1 from the previous studies (see \citealt{NEPW_catalog} for details), 2 from this study, and 9 indicates unavailability.}
    \tablenotetext{d}{The uncertainties of the photometric redshifts.}
\end{deluxetable*}

We finally construct the value-added NEPW band-merged catalog, including $z_{\rm phot, RF}$ and their uncertainties. The catalog includes 77755 sources with Subaru/HSC photometry because we use the colors from HSC photometry. Table \ref{tab2} summarizes the key characters of our value-added NEPW catalog. Table \ref{tab3} lists the individual objects in our value-added NEPW catalog.

\section{Discussion} \label{sec4}

\subsection{Significance of input features on the photometric redshift measurements} \label{sec4.1}

We investigate which photometric information in the input feature contributes significantly to the $z_{\rm phot, RF}$ measurements. The `permutation importance' is a widely used parameter to test the impact of certain input features in the machine learning prediction \citep{scikit-learn}. The permutation importance $I_j$ is defined as:
\begin{equation}
    I_j = s -\frac{1}{K}\sum^{K}_{k=1}s_{k,j},
\end{equation}
where $s$ is the RMS of the prediction based on all features, and $s_{k,j}$ is the RMS of the prediction based on the data sets with the randomly shuffled $j$-th feature.

\begin{figure}
    \centering
    \includegraphics[width=.5\textwidth]{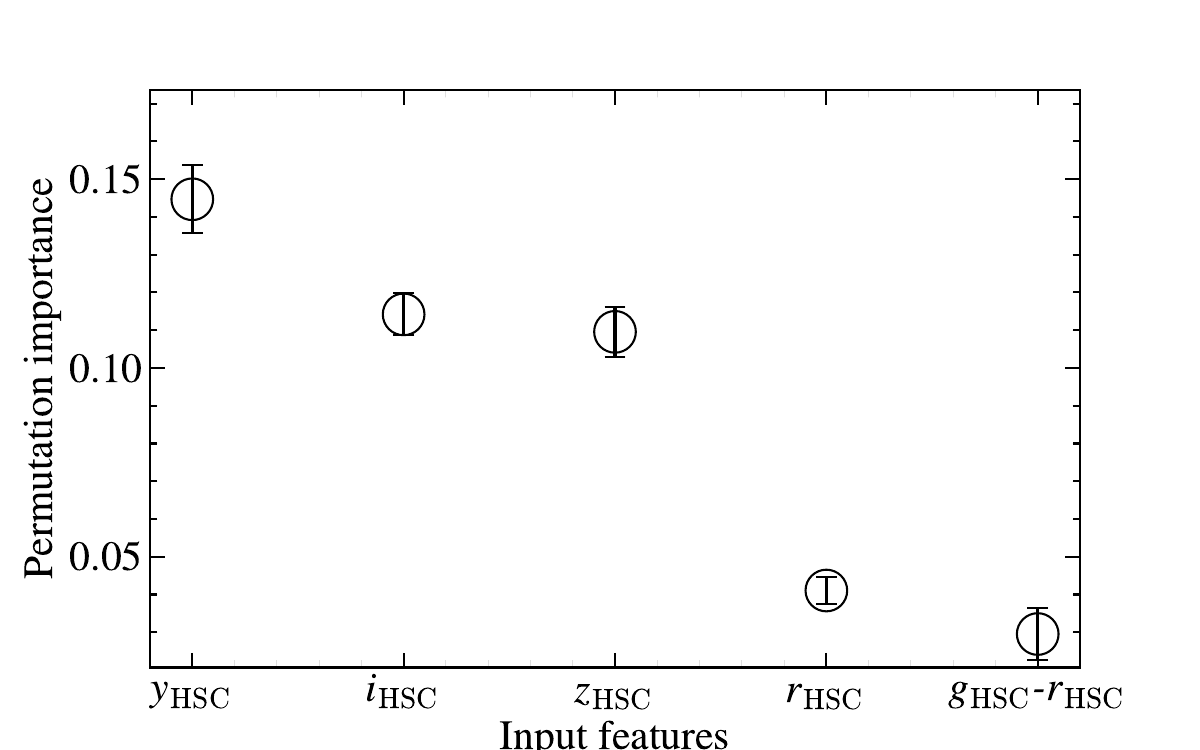}
    \caption{The permutation importance for top five input features in our $z_{\rm phot, RF}$ determination. The permutation importances for the other 51 input features are very small ($< 0.03$).}
    \label{fig8}
\end{figure}

Figure \ref{fig8} shows the result of the permutation importance computation for our 56 input features. There are five features with permutation importance larger than 0.03; $y_{\rm HSC},\ i_{\rm HSC},\ z_{\rm HSC}, r_{\rm HSC},\ g_{\rm HSC}-r_{\rm HSC}$. Not surprisingly, all top five features are from HSC photometry because all photometric sources we measured $z_{\rm phot, RF}$ have the HSC photometry. Because photometry from other observational data is incomplete for our input sources, the contribution of other band photometry is less significant in the determination of $z_{\rm phot, RF}$. 

\subsection{High fraction of catastrophic outliers for AGNs} \label{sec4.2}

While analyzing the result from the test set, we note that the catastrophic outlier rate is particularly high for AGNs ($\eta\mathrm{_{{AGN}}} \sim 40~\%$) compared to that for galaxies ($\eta\mathrm{_{galaxy}}<9$ \%). The high fraction of catastrophic outlier rate of AGNs originates from the lack of AGN samples in the training set; there are only 233 ($\sim 5~\%$) confirmed AGNs in our training set. Furthermore, the shapes of AGN SEDs are highly variable depending on the sources. Thus, estimating the $z_{\rm phot, RF}$ significantly suffers from the small size of the sample and high variability. Although the total AGN fraction in the NEPW catalog is low (234 AGAN candidates based on X-ray observations \citealp{Krumpe2015}), the $z_{\rm phot, RF}$ estimates for AGN candidates in the NEPW catalog require special caution. 

\subsection{Prediction accuracy depending on the input features} \label{sec4.3}

Our test with permutation importance indicates that the contribution of each input feature to the prediction can vary significantly. In other words, the inclusion/exclusion of certain parameters can affect the estimation of $z_{\rm phot, RF}$. Furthermore, we include the dummy values when some band photometry is not available. The impact of the choice of dummy values in the $z_{\rm phot, RF}$ needs to be investigated to test the reliability of $z_{\rm phot, RF}$. 

Here, we examine the accuracy of $z_{\rm phot, RF}$ measurement depending on the input features. We test the effect of the inclusion of magnitude uncertainties as input features. We also investigate the effect of dummy values we included to replace the missing photometry. The usage of color information in $z_{\rm phot, RF}$ is also examined. For this test, we generate the following datasets with various input features (summarized in Table \ref{tab4}):

\begin{deluxetable*}{cccc}
    \label{tab4}
    \tablecaption{Input datasets with various input features}
    \tablehead{\colhead{Set} & \colhead{magnitude uncertainty} & \colhead{missing photometry} & \colhead{included HSC photometry}}

    \startdata
    ${\rm Set_{fiducial}}$ & included & replaced with 1000 & mag., mag. uncertainties, and colors \\ \hline
    ${\rm Set_{no~err}}$ & excluded & replaced with 1000 & mag., mag. uncertainties, and colors \\
    ${\rm Set_{recon}}$ & excluded, reflected in mag. & replaced with 1000 & mag., mag. uncertainties, and colors \\ \hline
    ${\rm Set_{dum=-1000}}$ & included & replaced with -1000 & mag., mag. uncertainties, and colors\\
    ${\rm Set_{dum=-100}}$ & included & replaced with -100 & mag., mag. uncertainties, and colors \\ 
    ${\rm Set_{dum=100}}$ & included & replaced with 100 & mag., mag. uncertainties, and colors \\
    ${\rm Set_{interp}}$ & included & interpolated by adjacent photometry & mag., mag. uncertainties, and colors  \\ \hline
    ${\rm Set_{color~only}}$ & included & replaced with 1000 & mag. uncertainties, colors, and $r$-mag.\\
    ${\rm Set_{no color}}$ & included & replaced with 1000 & mag. and mag. uncertainties \\ \hline
    \enddata
\end{deluxetable*}

\begin{itemize}
\item ${\rm Set_{fiducial}}$ is a fiducial input dataset including 26 magnitudes and their uncertainties, and four HSC colors (see Section \ref{sec3.2}). 
\item ${\rm Set_{no~err}}$ is similar to Set$_{\rm fiducial}$, but without magnitude uncertainties. 
\item ${\rm Set_{recon}}$ is similar to Set$_{\rm no~err}$, but lists reconstructed magnitudes. Here, the reconstructed magnitudes indicate the sum of source magnitudes and their tentative magnitude uncertainties randomly extracted from the Gaussian distribution with the standard deviation corresponding to the photometric magnitude uncertainty. 
\item ${\rm Set_{dum = -1000}}$ is the same as Set$_{\rm fiducial}$, but the dummy value for missing photometry is -1000. 
\item ${\rm Set_{dum = -100}}$ is the same as Set$_{\rm fiducial}$, but the dummy value for missing photometry is -100. 
\item ${\rm Set_{dum = 100}}$ is the same as Set$_{\rm fiducial}$, but the dummy value for missing photometry is 100. 
\item ${\rm Set_{interp}}$ is similar to Set$_{\rm fiducial}$, but the missing photometry is replaced with the values derived from interpolation (or extrapolation) of nearby band photometry, rather than using dummy values.
\item ${\rm Set_{color~only}}$ contains 22 band photometry (and their uncertainties) except for HSC $gizY$-band photometry. The excluded HSC bands are replaced with HSC colors (i.e., \textit{g-r, r-i, i-z, z-Y}). 
\item ${\rm Set_{no color}}$ is similar to ${\rm Set_{fiducial}}$, but excludes Subaru/HSC colors (i.e., \textit{g-r, r-i, i-z, z-Y}) from the input features. 
\end{itemize}

\begin{figure*}[t]
    \centering
    \includegraphics[width=\textwidth]{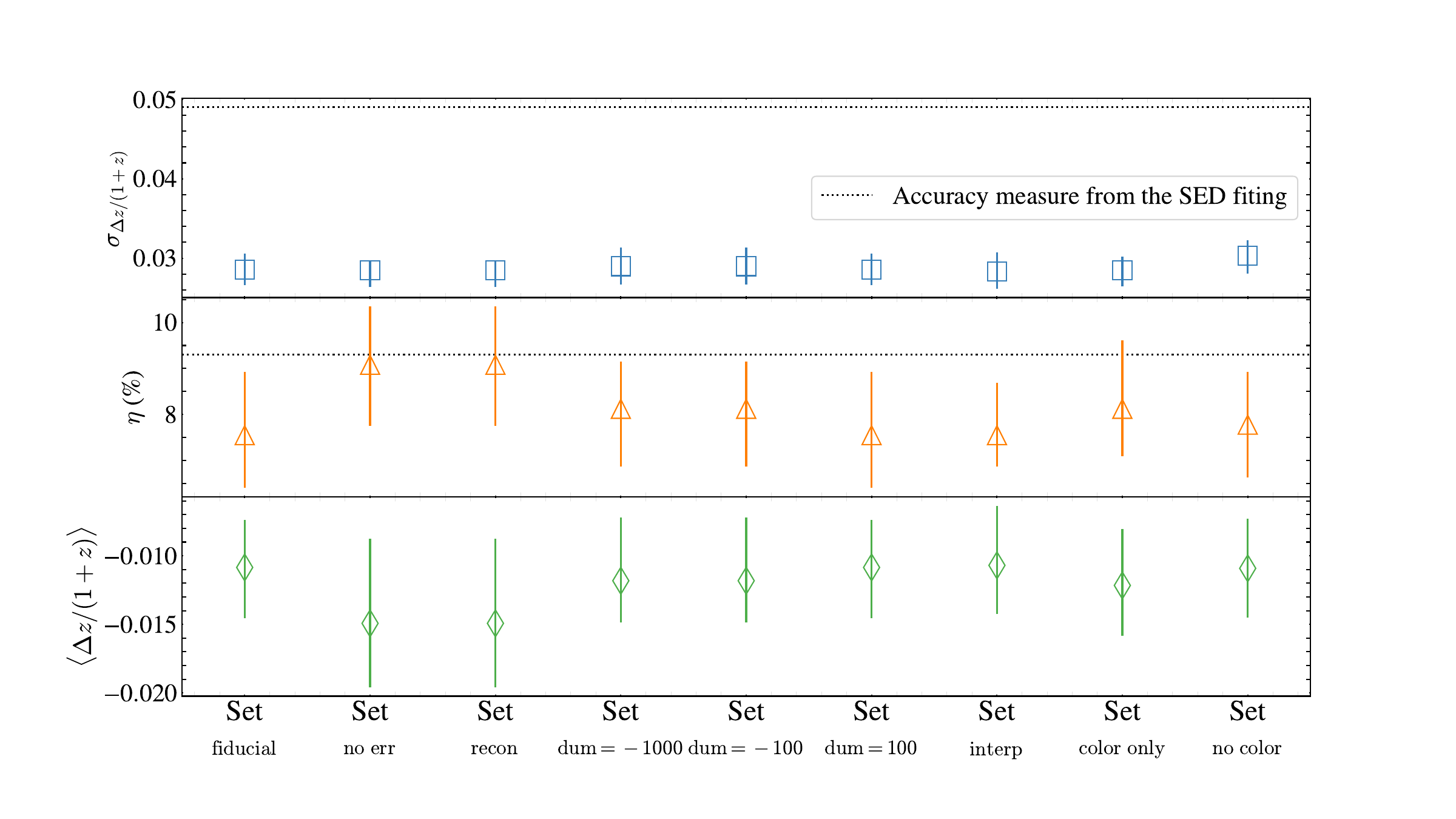}
    \caption{The accuracy metrics of the photometric redshift estimation depends on the input features. The top, middle, and bottom panels show the medians of $\sigma_{z/(1+z)}$, $\eta$, and $\langle \Delta z/(1+z) \rangle$ of each set with their $1\sigma$ scatter, respectively. The dotted lines in the top two panels indicate the accuracy metrics from the SED fit of \citet{Ho}.}
    \label{fig9}
\end{figure*}

Figure \ref{fig9} summarizes the accuracy metrics for $z_{\rm phot, RF}$ with respect to $z_{\rm spec}$ derived from nine different input datasets based on the 100 randomly selected training sets. The top panel shows the redshift difference dispersions (i.e., $\sigma_{\Delta z/(1+z)}$) of the nine datasets. The dispersions change little depending on the input datasets. Similarly, the catastrophic outlier fractions and the biases for the different input datasets are consistent with each other. Interestingly, the accuracy metrics (i.e., $\sigma_{\Delta z/(1+z)}$, $\eta$, and $ \langle\Delta z/(1+z)\rangle$) for $z_{\rm phot, RF}$ for all nine input datasets are lower than those for $z_{\rm phot, SED}$, suggesting the better performance of $z_{\rm phot, RF}$ estimations.

\subsection{Systematics at $z>0.8$} \label{sec4.4}

We note that the $z_{\rm phot, RF}$ are slightly lower than $z_{\rm spec}$ by $\sim 0.012$ across all input datasets. This redshift offset results from the high redshift ($z > 0.8$) targets. Discrepancies between $z_{\rm phot, RF}$ and $z_{\rm phot, SED}$ begin to appear at at $z_{\rm phot, RF}>0.8$ (Figure \ref{fig4}). The primary reason for the low accuracy of $z_{\rm phot, RF}$ at $z_{\rm phot, SED} > 0.8$ is the scarcity of training data at the redshift range. Because the random forest is only able to interpolate, the random forest precision is constrained by 1) the minimum and maximum redshifts and 2) the number density of redshifts in the training set. In our case, the training set only includes the objects with $z_{\rm spec} = [0., 1.758]$, and 8\% of them are at $z > 0.8$ (see also Figure \ref{fig2}). 

\begin{figure}
    \centering
    \includegraphics[width=.5\textwidth]{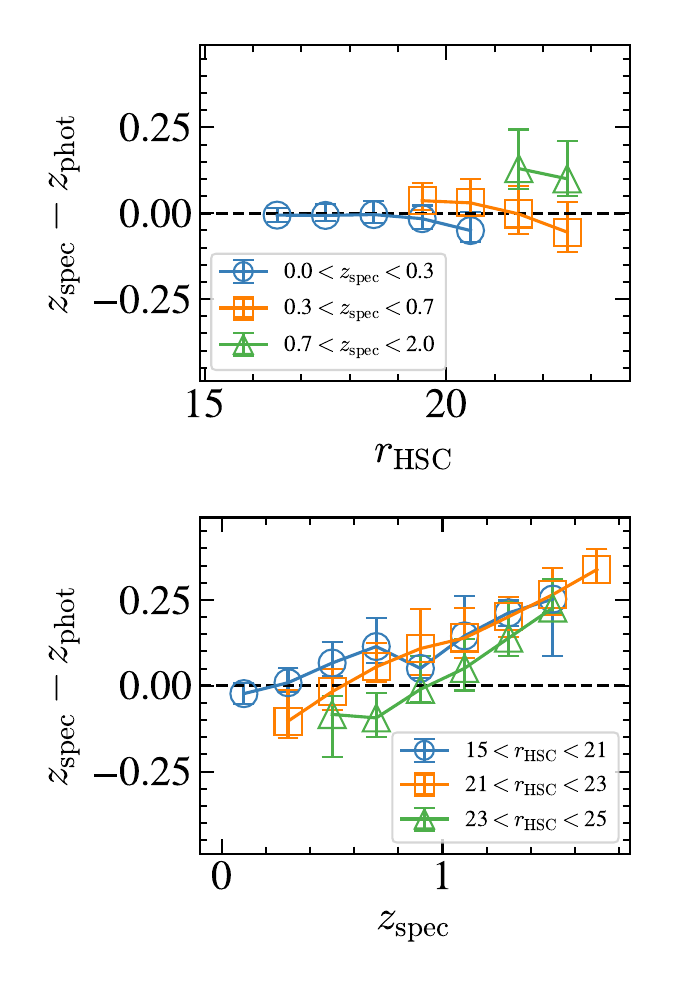}
    \caption{(Top) The redshift difference as a function of $r_{\rm HSC}$ in three $z_{\rm spec}$ bins. Blue circles, orange squares, and green triangles with error bars indicate the median redshift difference across $r_{\rm mag}$ with $0.0<z_{\rm spec}<0.4$, $0.4<z_{\rm spec}<0.8$, and $0.8<z_{\rm spec}<2.0$ with their $1\sigma$ distribution. (Bottom) The redshift difference as a function of $z_{\rm spec}$ in three $r_{\rm mag}$ bins. Blue circles, orange, squares, and green triangles indicate the median redshift difference across $z_{\rm spec}$ with $15<r_{\rm HSC}<21$, $21<r_{\rm HSC}<23$, and $23<r_{\rm HSC}<25$ with their $1\sigma$ distribution.}
    \label{fig10}
\end{figure}

In the top panel of Figure \ref{fig10}, we show the redshift difference as a function of HSC $r$-band magnitude for three redshift subsamples: $0.0<z<0.3$, $0.3<z<0.7$, and $0.7<z<2.0$. The redshift difference is more significant for the higher-redshift subsample, likely due to the smaller size of the training set for faint, high-redshift galaxies. Similarly, in the bottom panel of Figure \ref{fig10}, we display the redshift difference as a function of redshift for three HSC $r$-band magnitude subsamples. The redshift offset increases at higher redshift, reflecting the relative scarcity of high-$z$ objects in the training set.

\subsection{The accuracy of the photometric redshifts and the effect of incorrect photometry on the photometric redshift estimation}\label{sec4.5}

\begin{figure}[ht]
    \centering
    \includegraphics[width=.5\textwidth]{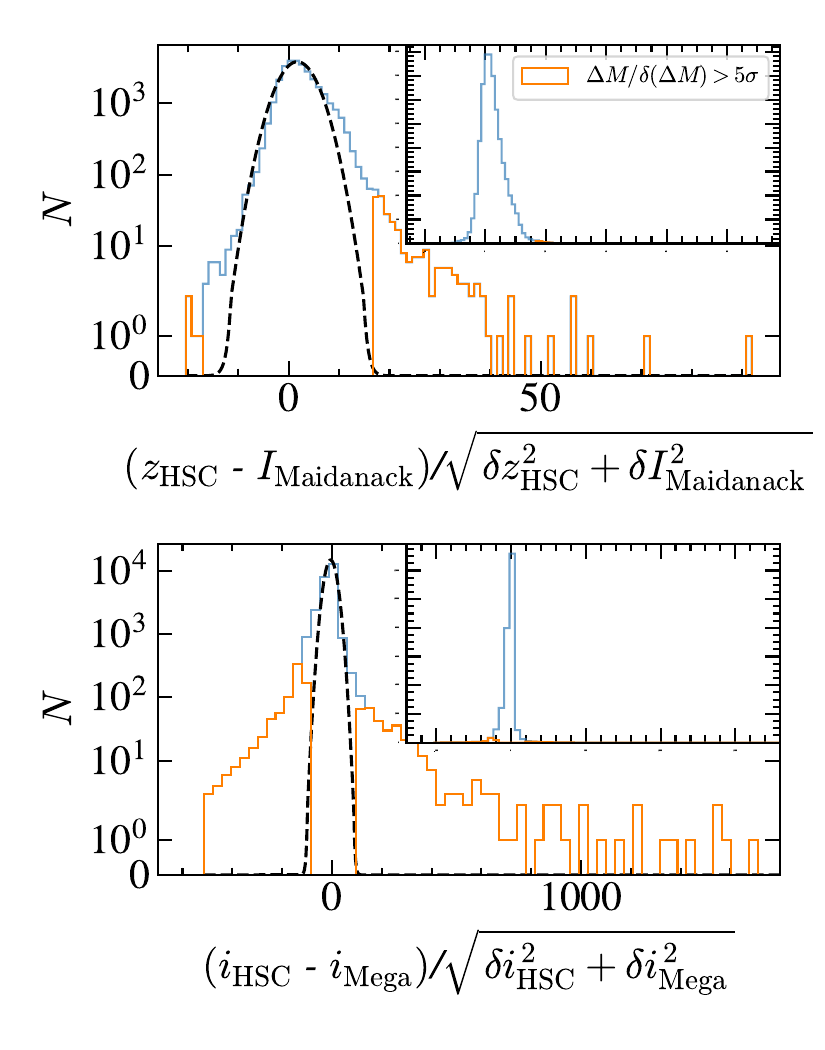}
    \caption{(Top) Distribution of the magnitude difference between HSC \textit{z} and Maidanak \textit{I} normalized by their uncertainties. The blue histogram shows the distribution of 30406 objects with Maidanak \textit{I}. The black dashed curve indicates the Gaussian best fit for the distribution. The orange histogram displays the distribution of $5\sigma$ outliers. The inset image shows the same distributions but in a linear scale for the number of objects.  
    (Bottom) Same as the top panel, but for HSC  \textit{i} and MegaCam \textit{i}.}
    \label{fig11}
\end{figure} 

Our photometric data are compiled from various photometric catalogs \citep{NEPW_catalog} obtained from multi-wavelength observations. In some regions within the NEPW field, there are photometric data from bands with similar wavelength ranges. For example, the Subaru/HSC \textit{z}-band with $\lambda_{\text{eff}}=8902$ \AA~ partially overlaps with the Maidanak/SNUCAM \textit{I}-band ($\lambda_{\text{eff}}=8603$ \AA). The Subaru/HSC \textit{i}-band with $\lambda_\text{eff}=7655$ \AA~ also overlaps with the CFHT/MegaCam \textit{i}-band ($\lambda_\text{eff}=7467$ \AA). \cite{Ho} showed that $\sim300$ objects in the NEPW catalog have magnitude differences larger than $5\sigma$ of the quadratic sum of the magnitude uncertainties in the two bands. The magnitude discrepancy is mainly due to the deblending of bright sources or sources with structures observed with instruments with different resolutions. The deep, high-resolution HSC images generally enable the identification of fainter sources or segments of bright sources. To avoid over-deblending in Subaru/HSC data, \citet{Ho} used only CFHT/MegaCAM or Maidanak/SNUCAM magnitudes for the SED fitting for the sources with large magnitude discrepancies.

Figure \ref{fig11} shows the differences between HSC \textit{z}-band and Maidanak \textit{I}-band magnitudes normalized by their uncertainties: $(z_{\rm HSC}$ - $I_{\rm Maidanak})$/$\sqrt{\delta z_{\rm HSC}^2+\delta I_{\rm Maidanak}^2}$ for the sources with overlapped photometry. The dashed line shows the best-fit Gaussian to the distribution. There are 262 sources with the normalized magnitude differences larger than $5\sigma$ (the orange histogram in Figure \ref{fig11}). Hereafter, we refer to these objects as `problematic' HSC photometry, because over-deblending in HSC photometry is the main cause. Similarly, there are 1131 sources identified as problematic due to magnitude differences greater than $5\sigma$ between HSC $i$-band and MegaCam $i$-band.

We test the impact of the magnitude discrepancy in the overlapped-band photometry on the $z_{\rm phot,RF}$ measurements. We prepare four input datasets that handle the problematic HSC photometry differently:
\begin{enumerate}
\item Set$_{\rm fiducial}$ is the input dataset including all magnitudes and their uncertainties from 26-bands (See Sections \ref{sec3.2}). 
\item Set 1 is the dataset including all magnitudes and their uncertainties from 26-bands. The dummy value of 1000 is used for the problematic HSC photometry. 
\item Set 2 is identical to Set 1, but we additionally replace the other band photometry with S/N$ < 3$ with the dummy value. 
\item Set 3 is similar to Set 1, but we also replace both Maidanak $I$ and MegaCam $i$ band magnitude and their uncertainties with the dummy value. 
\item Set 4 is identical to Set 3, but we additionally replace the other band photometry with S/N$ < 3$ with the dummy value. 
\end{enumerate}

\begin{figure*}[t]
    \centering
    \includegraphics[width=\textwidth]{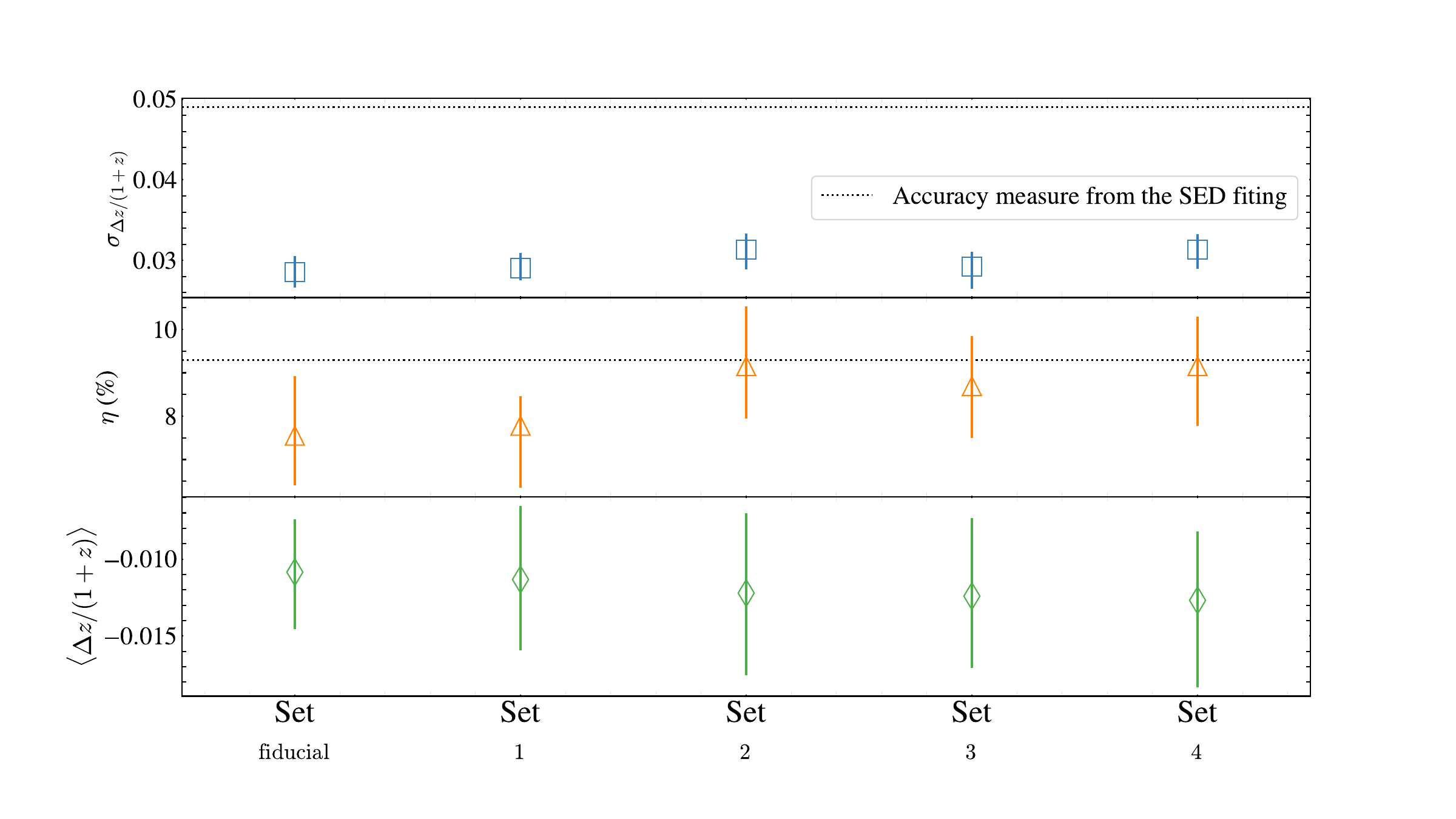}
    \caption{The accuracy metrics of the photometric redshift estimation for five different input datasets that handle the problematic HSC photometry, differently. The top, middle, and below panels show the medians of $\sigma_{z/(1+z)}$, $\eta$, and $\langle \Delta z/(1+z) \rangle$ of each set with their $1\sigma$ scatter, respectively. The dotted line indicates the accuracy metrics from the SED fit of \citet{Ho}.}
    \label{fig12}
\end{figure*}

We estimate $z_{\rm phot, RF}$ based on the four different input datasets and the Set$_{\rm fiducial}$. Figure \ref{fig12} displays the three accuracy metrics ($\sigma_{\Delta z/(1+z)}$, $\eta$, $\langle\Delta z/(1+z)\rangle$) derived from the different input datasets with respect to $z_{\rm spec}$. Horizontal lines show the accuracy metrics for $z_{\rm phot, SED}$ \citep{Ho}. The three accuracy metrics of $z_{\rm phot, RF}$ derived from five different datasets are all consistent with each other. In particular, the differences between Set$_{\rm fiducial}$ and other Sets are negligible. In other words, exclusion (or inclusion) of problematic HSC photometry has little effect on the $z_{\rm phot, RF}$ measurements. The test assures that $z_{\rm phot, RF}$ derived from Set$_{\rm fiducial}$, which we provided in the value-added catalog, are robust. 

\section{Conclusions} \label{sec5}

We derive photometric redshift of the sources in the NEPW field based on the extended multi-band photometry covering optical to infrared. We conduct our own MMT/Hectospec redshift survey to increase the number of spectroscopic redshifts. We then construct the random forest machine learning model to estimate the photometric redshifts trained with the spectroscopic sample based on spectroscopic redshifts we compile from literature and from our survey. We present the value-added NEPW catalog, including photometric redshifts ($z_{\rm phot, RF}$) for 77775 sources. 

\begin{itemize}

\item We obtain 2572 spectroscopic redshifts from our own MMT/Hectospec redshift survey for the NEPW field. We develop \texttt{RVSNUpy} (T. Kim et al. in preparation), a redshift measuring package based on a cross-correlation technique, to derive the spectroscopic redshifts from Hectospec spectra. Combined with the spectroscopic redshifts from the literature, we build the extended catalog, including 4421 spectroscopic redshifts at $z < 2$. 

\item We construct the random forest model to estimate photometric redshifts for the NEPW sources. We train the model based on the NEPW sample with 26-band photometry ranging $3000 < \lambda~({\rm \AA}) < 55000$ with the spectroscopic redshifts. The random forest model successfully measures photometric redshifts for 77755 sources within the NEPW field with a homogenous HSC photometry. 

\item We also devise a method to evaluate the uncertainty of random forest photometric redshift measurements. We found that there is a correlation between the $1\sigma$ of photometric and spectroscopic redshift difference (i.e., $\sigma_{spz}$) as a function of $\sigma_{\rm DT}$, the $1\sigma$ of the photometric redshift estimations from multiple decision trees for each object. We thus estimate the uncertainty of photometric redshift measurement for each object based on the $\sigma_{\rm DT}$ and the relation between $\sigma_{\rm DT}$ and $\sigma_{spz}$.  

\item We evaluate the quality of $z_{\rm phot, RF}$ by comparing it to $z_{\rm spec}$ and photometric redshifts derived from the SED fitting. The fraction of catastrophic outliers with $|(\Delta z)/(1+z)| > 0.15$ for our random forest measurements is comparable to that for the SED fitting measurements. However, $z_{\rm phot, RF}$ show a smaller scatter relative to $z_{\rm spec}$ than $z_{\rm phot, SED}$, indicating that $z_{\rm phot, RF}$ are generally closer to $z_{\rm spec}$ with smaller uncertainties.

\item We carefully examine $z_{\rm phot, RF}$ measurements based on the various input datasets to understand the impact of the inclusion of certain photometric bands and dummy values for missing photometry. Our test results assure that $z_{\rm phot, RF}$ measurements are robust regardless of the change in a number of input features. The missing photometry also has little impact on the $z_{\rm phot, RF}$ estimations. 

\item Finally, we present the value-added catalog for the NEPW band-merged catalog. We list the new spectroscopic redshifts from our Hectospec survey. In addition, we provide $z_{\rm phot, RF}$ we derive for 77755 sources based on our random forest machine learning model. 
    
\end{itemize}

Our data set will have an important legacy value for the studies near the NEP region, which will be targeted with space missions, including JWST \citep{2018PASP..130l4001J} and SPHEREx \citep{2016arXiv160607039D, 2018arXiv180505489D}.

\section*{Acknowledgements}
We thank the referee for the careful and insightful review of our manuscript. 
This work was supported by K-GMT Science Program (PID: UAO-G209-20B, UAO-G181-21B, UAO-G178-22A, SP ID: MMT-2020A-002, MMT-2020B-002, MMT-2021A-002, MMT-2021B-002) funded through Korean GMT Project operated by Korea Astronomy and Space Science Institute (KASI).
N.H. and B.-G. P. acknowledge the support by the Korea Astronomy and Space Science Institute (KASI) grant funded by the Korean government (MSIT; No. 2023-1-860-02, International Optical Observatory Project). 
HSH acknowledges the support by Samsung Electronic Co., Ltd. (Project Number IO220811-01945-01), the National Research Foundation of Korea (NRF) grant funded by the Korea government (MSIT), NRF-2021R1A2C1094577, and Hyunsong Educational \& Cultural Foundation.
HS acknowledges the support by the National Research Foundation of Korea(NRF) grant funded by the Korea government(MSIT) (2022M3K3A1093827).
SH acknowledges the support of The Australian Research Council Centre of Excellence for Gravitational Wave Discovery (OzGrav) and the Australian Research Council Centre of Excellence for All Sky Astrophysics in 3 Dimensions (ASTRO 3D), through project number CE17010000 and CE170100013, respectively.
JS acknowledges the support by the National Research Foundation of Korea (NRF) grant funded by the Korea government (MSIT) (RS-2023-00210597).
This work is based on observations with
AKARI, a JAXA project with the participation of ESA, universities,
and companies in Japan, Korea, and the UK.

\clearpage

\bibliography{ms}

\begin{thebibliography}{}
\expandafter\ifx\csname natexlab\endcsname\relax\def\natexlab#1{#1}\fi
\providecommand{\url}[1]{\href{#1}{#1}}
\providecommand{\dodoi}[1]{doi:~\href{http://doi.org/#1}{\nolinkurl{#1}}}
\providecommand{\doeprint}[1]{\href{http://ascl.net/#1}{\nolinkurl{http://ascl.net/#1}}}
\providecommand{\doarXiv}[1]{\href{https://arxiv.org/abs/#1}{\nolinkurl{https://arxiv.org/abs/#1}}}

\bibitem[{{Ahumada} {et~al.}(2020){Ahumada}, {Allende Prieto}, {Almeida},
  {Anders}, {Anderson}, {Andrews}, {Anguiano}, {Arcodia}, {Armengaud},
  {Aubert}, {Avila}, {Avila-Reese}, {Badenes}, {Balland}, {Barger},
  {Barrera-Ballesteros}, {Basu}, {Bautista}, {Beaton}, {Beers}, {Benavides},
  {Bender}, {Bernardi}, {Bershady}, {Beutler}, {Bidin}, {Bird}, {Bizyaev},
  {Blanc}, {Blanton}, {Boquien}, {Borissova}, {Bovy}, {Brandt}, {Brinkmann},
  {Brownstein}, {Bundy}, {Bureau}, {Burgasser}, {Burtin}, {Cano-D{\'\i}az},
  {Capasso}, {Cappellari}, {Carrera}, {Chabanier}, {Chaplin}, {Chapman},
  {Cherinka}, {Chiappini}, {Doohyun Choi}, {Chojnowski}, {Chung}, {Clerc},
  {Coffey}, {Comerford}, {Comparat}, {da Costa}, {Cousinou}, {Covey}, {Crane},
  {Cunha}, {Ilha}, {Dai}, {Damsted}, {Darling}, {Davidson}, {Davies}, {Dawson},
  {De}, {de la Macorra}, {De Lee}, {Queiroz}, {Deconto Machado}, {de la Torre},
  {Dell'Agli}, {du Mas des Bourboux}, {Diamond-Stanic}, {Dillon}, {Donor},
  {Drory}, {Duckworth}, {Dwelly}, {Ebelke}, {Eftekharzadeh}, {Davis Eigenbrot},
  {Elsworth}, {Eracleous}, {Erfanianfar}, {Escoffier}, {Fan}, {Farr},
  {Fern{\'a}ndez-Trincado}, {Feuillet}, {Finoguenov}, {Fofie},
  {Fraser-McKelvie}, {Frinchaboy}, {Fromenteau}, {Fu}, {Galbany}, {Garcia},
  {Garc{\'\i}a-Hern{\'a}ndez}, {Garma Oehmichen}, {Ge}, {Geimba Maia},
  {Geisler}, {Gelfand}, {Goddy}, {Gonzalez-Perez}, {Grabowski}, {Green},
  {Grier}, {Guo}, {Guy}, {Harding}, {Hasselquist}, {Hawken}, {Hayes}, {Hearty},
  {Hekker}, {Hogg}, {Holtzman}, {Horta}, {Hou}, {Hsieh}, {Huber}, {Hunt}, {Ider
  Chitham}, {Imig}, {Jaber}, {Jimenez Angel}, {Johnson}, {Jones},
  {J{\"o}nsson}, {Jullo}, {Kim}, {Kinemuchi}, {Kirkpatrick}, {Kite}, {Klaene},
  {Kneib}, {Kollmeier}, {Kong}, {Kounkel}, {Krishnarao}, {Lacerna}, {Lan},
  {Lane}, {Law}, {Le Goff}, {Leung}, {Lewis}, {Li}, {Lian}, {Lin}, {Long},
  {Longa-Pe{\~n}a}, {Lundgren}, {Lyke}, {Mackereth}, {MacLeod}, {Majewski},
  {Manchado}, {Maraston}, {Martini}, {Masseron}, {Masters}, {Mathur},
  {McDermid}, {Merloni}, {Merrifield}, {M{\'e}sz{\'a}ros}, {Miglio}, {Minniti},
  {Minsley}, {Miyaji}, {Mohammad}, {Mosser}, {Mueller}, {Muna},
  {Mu{\~n}oz-Guti{\'e}rrez}, {Myers}, {Nadathur}, {Nair}, {Nandra}, {Correa do
  Nascimento}, {Nevin}, {Newman}, {Nidever}, {Nitschelm}, {Noterdaeme},
  {O'Connell}, {Olmstead}, {Oravetz}, {Oravetz}, {Osorio}, {Pace}, {Padilla},
  {Palanque-Delabrouille}, {Palicio}, {Pan}, {Pan}, {Parker}, {Paviot},
  {Peirani}, {Ram{\'r}ez}, {Penny}, {Percival}, {Perez-Fournon},
  {P{\'e}rez-R{\`a}fols}, {Petitjean}, {Pieri}, {Pinsonneault}, {Poovelil},
  {Povick}, {Prakash}, {Price-Whelan}, {Raddick}, {Raichoor}, {Ray}, {Rembold},
  {Rezaie}, {Riffel}, {Riffel}, {Rix}, {Robin}, {Roman-Lopes},
  {Rom{\'a}n-Z{\'u}{\~n}iga}, {Rose}, {Ross}, {Rossi}, {Rowlands}, {Rubin},
  {Salvato}, {S{\'a}nchez}, {S{\'a}nchez-Menguiano}, {S{\'a}nchez-Gallego},
  {Sayres}, {Schaefer}, {Schiavon}, {Schimoia}, {Schlafly}, {Schlegel},
  {Schneider}, {Schultheis}, {Schwope}, {Seo}, {Serenelli}, {Shafieloo},
  {Shamsi}, {Shao}, {Shen}, {Shetrone}, {Shirley}, {Silva Aguirre}, {Simon},
  {Skrutskie}, {Slosar}, {Smethurst}, {Sobeck}, {Sodi}, {Souto}, {Stark},
  {Stassun}, {Steinmetz}, {Stello}, {Stermer}, {Storchi-Bergmann},
  {Streblyanska}, {Stringfellow}, {Stutz}, {Su{\'a}rez}, {Sun},
  {Taghizadeh-Popp}, {Talbot}, {Tayar}, {Thakar}, {Theriault}, {Thomas},
  {Thomas}, {Tinker}, {Tojeiro}, {Toledo}, {Tremonti}, {Troup}, {Tuttle},
  {Unda-Sanzana}, {Valentini}, {Vargas-Gonz{\'a}lez}, {Vargas-Maga{\~n}a},
  {V{\'a}zquez-Mata}, {Vivek}, {Wake}, {Wang}, {Weaver}, {Weijmans}, {Wild},
  {Wilson}, {Wilson}, {Wolthuis}, {Wood-Vasey}, {Yan}, {Yang}, {Y{\`e}che},
  {Zamora}, {Zarrouk}, {Zasowski}, {Zhang}, {Zhao}, {Zhao}, {Zheng}, {Zheng},
  {Zhu}, \& {Zou}}]{SDSS_dr16}
{Ahumada}, R., {Allende Prieto}, C., {Almeida}, A., {et~al.} 2020, \apjs, 249,
  3, \dodoi{10.3847/1538-4365/ab929e}

\bibitem[{{Almeida} {et~al.}(2023){Almeida}, {Anderson},
  {Argudo-Fern{\'a}ndez}, {Badenes}, {Barger}, {Barrera-Ballesteros}, {Bender},
  {Benitez}, {Besser}, {Bird}, {Bizyaev}, {Blanton}, {Bochanski}, {Bovy},
  {Brandt}, {Brownstein}, {Buchner}, {Bulbul}, {Burchett}, {Cano D{\'\i}az},
  {Carlberg}, {Casey}, {Chandra}, {Cherinka}, {Chiappini}, {Coker}, {Comparat},
  {Conroy}, {Contardo}, {Cortes}, {Covey}, {Crane}, {Cunha}, {Dabbieri},
  {Davidson}, {Davis}, {de Andrade Queiroz}, {De Lee}, {M{\'e}ndez Delgado},
  {Demasi}, {Di Mille}, {Donor}, {Dow}, {Dwelly}, {Eracleous}, {Eriksen},
  {Fan}, {Farr}, {Frederick}, {Fries}, {Frinchaboy}, {G{\"a}nsicke}, {Ge},
  {Gonz{\'a}lez {\'A}vila}, {Grabowski}, {Grier}, {Guiglion}, {Gupta}, {Hall},
  {Hawkins}, {Hayes}, {Hermes}, {Hern{\'a}ndez-Garc{\'\i}a}, {Hogg},
  {Holtzman}, {Ibarra-Medel}, {Ji}, {Jofre}, {Johnson}, {Jones}, {Kinemuchi},
  {Kluge}, {Koekemoer}, {Kollmeier}, {Kounkel}, {Krishnarao}, {Krumpe},
  {Lacerna}, {Lago}, {Laporte}, {Liu}, {Liu}, {Liu}, {Lopes}, {Macktoobian},
  {Majewski}, {Malanushenko}, {Maoz}, {Masseron}, {Masters}, {Matijevic},
  {McBride}, {Medan}, {Merloni}, {Morrison}, {Myers}, {M{\'e}sz{\'a}ros},
  {Negrete}, {Nidever}, {Nitschelm}, {Oravetz}, {Oravetz}, {Pan}, {Peng},
  {Pinsonneault}, {Pogge}, {Qiu}, {Ramirez}, {Rix}, {Fern{\'a}ndez Rosso},
  {Runnoe}, {Salvato}, {Sanchez}, {Santana}, {Saydjari}, {Sayres},
  {Schlaufman}, {Schneider}, {Schwope}, {Serna}, {Shen}, {Sobeck}, {Song},
  {Souto}, {Spoo}, {Stassun}, {Steinmetz}, {Straumit}, {Stringfellow},
  {S{\'a}nchez-Gallego}, {Taghizadeh-Popp}, {Tayar}, {Thakar}, {Tissera},
  {Tkachenko}, {Hernandez Toledo}, {Trakhtenbrot}, {Fern{\'a}ndez-Trincado},
  {Troup}, {Trump}, {Tuttle}, {Ulloa}, {Vazquez-Mata}, {Vera Alfaro},
  {Villanova}, {Wachter}, {Weijmans}, {Wheeler}, {Wilson}, {Wojno}, {Wolf},
  {Xue}, {Ybarra}, {Zari}, \& {Zasowski}}]{sdss}
{Almeida}, A., {Anderson}, S.~F., {Argudo-Fern{\'a}ndez}, M., {et~al.} 2023,
  \apjs, 267, 44, \dodoi{10.3847/1538-4365/acda98}

\bibitem[{{Arnouts} {et~al.}(1999){Arnouts}, {Cristiani}, {Moscardini},
  {Matarrese}, {Lucchin}, {Fontana}, \& {Giallongo}}]{Arnouts99}
{Arnouts}, S., {Cristiani}, S., {Moscardini}, L., {et~al.} 1999, \mnras, 310,
  540, \dodoi{10.1046/j.1365-8711.1999.02978.x}

\bibitem[{{Barrufet} {et~al.}(2020){Barrufet}, {Pearson}, {Serjeant},
  {Ma{\l}ek}, {Baronchelli}, {Campos-Varillas}, {White}, {Valtchanov},
  {Matsuhara}, {Conversi}, {Kim}, {Goto}, {Oi}, {Malkan}, {Kim}, {Ikeda},
  {Takagi}, {Toba}, \& {Miyaji}}]{Barrufet20}
{Barrufet}, L., {Pearson}, C., {Serjeant}, S., {et~al.} 2020, \aap, 641, A129,
  \dodoi{10.1051/0004-6361/202037838}

\bibitem[{{Baum}(1957)}]{Baum1957}
{Baum}, W.~A. 1957, \aj, 62, 6, \dodoi{10.1086/107433}

\bibitem[{{Ben{\'\i}tez}(2000)}]{Benitz00}
{Ben{\'\i}tez}, N. 2000, \apj, 536, 571, \dodoi{10.1086/308947}

\bibitem[{{Bolzonella} {et~al.}(2000){Bolzonella}, {Miralles}, \&
  {Pell{\'o}}}]{2000A&A...363..476B}
{Bolzonella}, M., {Miralles}, J.~M., \& {Pell{\'o}}, R. 2000, \aap, 363, 476,
  \dodoi{10.48550/arXiv.astro-ph/0003380}

\bibitem[{{Breiman}(2001)}]{random-forest}
{Breiman}, L. 2001, Machine Learning, 45, 5, \dodoi{10.1023/A:1010933404324}

\bibitem[{{Cavuoti} {et~al.}(2012){Cavuoti}, {Brescia}, {Longo}, \&
  {Mercurio}}]{2012A&A...546A..13C}
{Cavuoti}, S., {Brescia}, M., {Longo}, G., \& {Mercurio}, A. 2012, \aap, 546,
  A13, \dodoi{10.1051/0004-6361/201219755}

\bibitem[{{Chen} {et~al.}(2021){Chen}, {Goto}, {Kim}, {Wang}, {Santos}, {Ho},
  {Hashimoto}, {Poliszczuk}, {Pollo}, {Trippe}, {Miyaji}, {Toba}, {Malkan},
  {Serjeant}, {Pearson}, {Hwang}, {Kim}, {Shim}, {Lu}, {Hsiao}, {Huang},
  {Herrera-Endoqui}, {Bravo-Navarro}, \& {Matsuhara}}]{chen21}
{Chen}, B.~H., {Goto}, T., {Kim}, S.~J., {et~al.} 2021, \mnras, 501, 3951,
  \dodoi{10.1093/mnras/staa3865}

\bibitem[{{Chiang} {et~al.}(2019){Chiang}, {Goto}, {Hashimoto}, {Kim},
  {Matsuhara}, \& {Oi}}]{Chiang19}
{Chiang}, C.-Y., {Goto}, T., {Hashimoto}, T., {et~al.} 2019, \pasj, 71, 31,
  \dodoi{10.1093/pasj/psz012}

\bibitem[{{Cohen} \& {Cohen}(1975)}]{dummyvalue}
{Cohen}, J., \& {Cohen}, P. 1975, Applied multiple regression/correlation
  analysis for the behavioral sciences. (Lawrence Erlbaum)

\bibitem[{{Davis} \& {Peebles}(1983)}]{intro1}
{Davis}, M., \& {Peebles}, P.~J.~E. 1983, \apj, 267, 465,
  \dodoi{10.1086/160884}

\bibitem[{{DESI Collaboration} {et~al.}(2024){DESI Collaboration}, {Adame},
  {Aguilar}, {Ahlen}, {Alam}, {Aldering}, {Alexander}, {Alfarsy}, {Allende
  Prieto}, {Alvarez}, {Alves}, {Anand}, {Andrade-Oliveira}, {Armengaud},
  {Asorey}, {Avila}, {Aviles}, {Bailey}, {Balaguera-Antol{\'\i}nez},
  {Ballester}, {Baltay}, {Bault}, {Bautista}, {Behera}, {Beltran}, {BenZvi},
  {Beraldo e Silva}, {Bermejo-Climent}, {Berti}, {Besuner}, {Beutler},
  {Bianchi}, {Blake}, {Blum}, {Bolton}, {Brieden}, {Brodzeller}, {Brooks},
  {Brown}, {Buckley-Geer}, {Burtin}, {Cabayol-Garcia}, {Cai}, {Canning},
  {Cardiel-Sas}, {Carnero Rosell}, {Castander}, {Cervantes-Cota}, {Chabanier},
  {Chaussidon}, {Chaves-Montero}, {Chen}, {Chen}, {Chuang}, {Claybaugh},
  {Cole}, {Cooper}, {Cuceu}, {Davis}, {Dawson}, {de Belsunce}, {de la Cruz},
  {de la Macorra}, {Della Costa}, {de Mattia}, {Demina}, {Demirbozan},
  {DeRose}, {Dey}, {Dey}, {Dhungana}, {Ding}, {Ding}, {Doel}, {Doshi},
  {Douglass}, {Edge}, {Eftekharzadeh}, {Eisenstein}, {Elliott}, {Ereza},
  {Escoffier}, {Fagrelius}, {Fan}, {Fanning}, {Fawcett}, {Ferraro}, {Flaugher},
  {Font-Ribera}, {Forero-Romero}, {Forero-S{\'a}nchez}, {Frenk},
  {G{\"a}nsicke}, {Garc{\'\i}a}, {Garc{\'\i}a-Bellido}, {Garcia-Quintero},
  {Garrison}, {Gil-Mar{\'\i}n}, {Golden-Marx}, {Gontcho A Gontcho},
  {Gonzalez-Morales}, {Gonzalez-Perez}, {Gordon}, {Graur}, {Green}, {Gruen},
  {Guy}, {Hadzhiyska}, {Hahn}, {Han}, {Hanif}, {Herrera-Alcantar}, {Honscheid},
  {Hou}, {Howlett}, {Huterer}, {Ir{\v{s}}i{\v{c}}}, {Ishak}, {Jacques}, {Jana},
  {Jiang}, {Jimenez}, {Jing}, {Joudaki}, {Joyce}, {Jullo}, {Juneau},
  {Kara{\c{c}}ayl{\i}}, {Karim}, {Kehoe}, {Kent}, {Khederlarian}, {Kim},
  {Kirkby}, {Kisner}, {Kitaura}, {Kizhuprakkat}, {Kneib}, {Koposov},
  {Kov{\'a}cs}, {Kremin}, {Krolewski}, {L'Huillier}, {Lahav}, {Lambert},
  {Lamman}, {Lan}, {Landriau}, {Lang}, {Lange}, {Lasker}, {Leauthaud}, {Le
  Guillou}, {Levi}, {Li}, {Linder}, {Lyons}, {Magneville}, {Manera}, {Manser},
  {Margala}, {Martini}, {McDonald}, {Medina}, {Medina-Varela}, {Meisner},
  {Mena-Fern{\'a}ndez}, {Meneses-Rizo}, {Mezcua}, {Miquel}, {Montero-Camacho},
  {Moon}, {Moore}, {Moustakas}, {Mueller}, {Mundet}, {Mu{\~n}oz-Guti{\'e}rrez},
  {Myers}, {Nadathur}, {Napolitano}, {Neveux}, {Newman}, {Nie}, {Nikutta},
  {Niz}, {Norberg}, {Noriega}, {Paillas}, {Palanque-Delabrouille}, {Palmese},
  {Pan}, {Parkinson}, {Penmetsa}, {Percival}, {P{\'e}rez-Fern{\'a}ndez},
  {P{\'e}rez-R{\`a}fols}, {Pieri}, {Poppett}, {Porredon}, {Pothier}, {Prada},
  {Pucha}, {Raichoor}, {Ram{\'\i}rez-P{\'e}rez}, {Ramirez-Solano},
  {Rashkovetskyi}, {Ravoux}, {Rocher}, {Rockosi}, {Ross}, {Rossi}, {Ruggeri},
  {Ruhlmann-Kleider}, {Sabiu}, {Said}, {Saintonge}, {Samushia}, {Sanchez},
  {Saulder}, {Schaan}, {Schlafly}, {Schlegel}, {Scholte}, {Schubnell}, {Seo},
  {Shafieloo}, {Sharples}, {Sheu}, {Silber}, {Sinigaglia}, {Siudek}, {Slepian},
  {Smith}, {Soumagnac}, {Sprayberry}, {Stephey}, {Su{\'a}rez-P{\'e}rez}, {Sun},
  {Tan}, {Tarl{\'e}}, {Tojeiro}, {Ure{\~n}a-L{\'o}pez}, {Vaisakh}, {Valcin},
  {Valdes}, {Valluri}, {Vargas-Maga{\~n}a}, {Variu}, {Verde}, {Walther},
  {Wang}, {Wang}, {Weaver}, {Weaverdyck}, {Wechsler}, {White}, {Xie}, {Yang},
  {Y{\`e}che}, {Yu}, {Yuan}, {Zhang}, {Zhang}, {Zhao}, {Zheng}, {Zhou}, {Zhou},
  {Zou}, {Zou}, \& {Zu}}]{desi}
{DESI Collaboration}, {Adame}, A.~G., {Aguilar}, J., {et~al.} 2024, \aj, 168,
  58, \dodoi{10.3847/1538-3881/ad3217}

\bibitem[{{D{\'\i}az Tello} {et~al.}(2017){D{\'\i}az Tello}, {Miyaji},
  {Ishigaki}, {Krumpe}, {Ueda}, {Brunner}, {Goto}, {Hanami}, \&
  {Toba}}]{specz5}
{D{\'\i}az Tello}, J., {Miyaji}, T., {Ishigaki}, T., {et~al.} 2017, \aap, 604,
  A14, \dodoi{10.1051/0004-6361/201730611}

\bibitem[{{D'Isanto} \& {Polsterer}(2018)}]{D'Isanto2018}
{D'Isanto}, A., \& {Polsterer}, K.~L. 2018, \aap, 609, A111,
  \dodoi{10.1051/0004-6361/201731326}

\bibitem[{{Dor{\'e}} {et~al.}(2016){Dor{\'e}}, {Werner}, {Ashby}, {Banerjee},
  {Battaglia}, {Bauer}, {Benjamin}, {Bleem}, {Bock}, {Boogert}, {Bull},
  {Capak}, {Chang}, {Chiar}, {Cohen}, {Cooray}, {Crill}, {Cushing}, {de
  Putter}, {Driver}, {Eifler}, {Feng}, {Ferraro}, {Finkbeiner}, {Gaudi},
  {Greene}, {Hillenbrand}, {H{\"o}flich}, {Hsiao}, {Huffenberger}, {Jansen},
  {Jeong}, {Joshi}, {Kim}, {Kim}, {Kirkpatrick}, {Korngut}, {Krause}, {Kriek},
  {Leistedt}, {Li}, {Lisse}, {Mauskopf}, {Mechtley}, {Melnick}, {Mohr},
  {Murphy}, {Neben}, {Neufeld}, {Nguyen}, {Pierpaoli}, {Pyo}, {Rhodes},
  {Sandstrom}, {Schaan}, {Schlaufman}, {Silverman}, {Su}, {Stassun}, {Stevens},
  {Strauss}, {Tielens}, {Tsai}, {Tolls}, {Unwin}, {Viero}, {Windhorst}, \&
  {Zemcov}}]{2016arXiv160607039D}
{Dor{\'e}}, O., {Werner}, M.~W., {Ashby}, M., {et~al.} 2016, arXiv e-prints,
  arXiv:1606.07039, \dodoi{10.48550/arXiv.1606.07039}

\bibitem[{{Dor{\'e}} {et~al.}(2018){Dor{\'e}}, {Werner}, {Ashby}, {Bleem},
  {Bock}, {Burt}, {Capak}, {Chang}, {Chaves-Montero}, {Chen}, {Civano},
  {Cleeves}, {Cooray}, {Crill}, {Crossfield}, {Cushing}, {de la Torre},
  {DiMatteo}, {Dvory}, {Dvorkin}, {Espaillat}, {Ferraro}, {Finkbeiner},
  {Greene}, {Hewitt}, {Hogg}, {Huffenberger}, {Jun}, {Ilbert}, {Jeong},
  {Johnson}, {Kim}, {Kirkpatrick}, {Kowalski}, {Korngut}, {Li}, {Lisse},
  {MacGregor}, {Mamajek}, {Mauskopf}, {Melnick}, {M{\'e}nard}, {Neyrinck},
  {{\"O}berg}, {Pisani}, {Rocca}, {Salvato}, {Schaan}, {Scoville}, {Song},
  {Stevens}, {Tenneti}, {Teplitz}, {Tolls}, {Unwin}, {Urry}, {Wandelt},
  {Williams}, {Wilner}, {Windhorst}, {Wolk}, {Yorke}, \&
  {Zemcov}}]{2018arXiv180505489D}
{Dor{\'e}}, O., {Werner}, M.~W., {Ashby}, M. L.~N., {et~al.} 2018, arXiv
  e-prints, arXiv:1805.05489, \dodoi{10.48550/arXiv.1805.05489}

\bibitem[{{Eriksen} {et~al.}(2019){Eriksen}, {Alarcon}, {Gaztanaga}, {Amara},
  {Cabayol}, {Carretero}, {Castander}, {Crocce}, {Delfino}, {De Vicente},
  {Fernandez}, {Fosalba}, {Garcia-Bellido}, {Hildebrandt}, {Hoekstra},
  {Joachimi}, {Norberg}, {Miquel}, {Padilla}, {Refregier}, {Sanchez},
  {Serrano}, {Sevilla-Noarbe}, {Tallada}, {Tonello}, \&
  {Tortorelli}}]{Eriksen2019}
{Eriksen}, M., {Alarcon}, A., {Gaztanaga}, E., {et~al.} 2019, \mnras, 484,
  4200, \dodoi{10.1093/mnras/stz204}

\bibitem[{{Fabricant} {et~al.}(2005){Fabricant}, {Fata}, {Roll}, {Hertz},
  {Caldwell}, {Gauron}, {Geary}, {McLeod}, {Szentgyorgyi}, {Zajac}, {Kurtz},
  {Barberis}, {Bergner}, {Brown}, {Conroy}, {Eng}, {Geller}, {Goddard},
  {Honsa}, {Mueller}, {Mink}, {Ordway}, {Tokarz}, {Woods}, {Wyatt}, {Epps}, \&
  {Dell'Antonio}}]{hectospec}
{Fabricant}, D., {Fata}, R., {Roll}, J., {et~al.} 2005, \pasp, 117, 1411,
  \dodoi{10.1086/497385}

\bibitem[{{Goto} {et~al.}(2017){Goto}, {Toba}, {Utsumi}, {Oi}, {Takagi},
  {Malkan}, {Ohayma}, {Murata}, {Price}, {Karouzos}, {Matsuhara}, {Nakagawa},
  {Wada}, {Serjeant}, {Burgarella}, {Buat}, {Takada}, {Miyazaki}, {Oguri},
  {Miyaji}, {Oyabu}, {White}, {Takeuchi}, {Inami}, {Perason}, {Malek},
  {Marchetti}, {Lee}, {Im}, {Kim}, {Koptelova}, {Chao}, {Wu}, {AKARI NEP Survey
  Team}, \& {AKARI All Sky Survey Team}}]{deep_hsc1}
{Goto}, T., {Toba}, Y., {Utsumi}, Y., {et~al.} 2017, Publication of Korean
  Astronomical Society, 32, 225, \dodoi{10.5303/PKAS.2017.32.1.225}

\bibitem[{{Goto} {et~al.}(2019){Goto}, {Oi}, {Utsumi}, {Momose}, {Matsuhara},
  {Hashimoto}, {Toba}, {Ohyama}, {Takagi}, {Chiang}, {Kim}, {Kilerci Eser},
  {Malkan}, {Kim}, {Miyaji}, {Im}, {Nakagawa}, {Jeong}, {Pearson}, {Barrufet},
  {Sedgwick}, {Burgarella}, {Buat}, \& {Ikeda}}]{goto19}
{Goto}, T., {Oi}, N., {Utsumi}, Y., {et~al.} 2019, \pasj, 71, 30,
  \dodoi{10.1093/pasj/psz009}

\bibitem[{{Henghes} {et~al.}(2022){Henghes}, {Thiyagalingam}, {Pettitt}, {Hey},
  \& {Lahav}}]{Henghes2022}
{Henghes}, B., {Thiyagalingam}, J., {Pettitt}, C., {Hey}, T., \& {Lahav}, O.
  2022, \mnras, 512, 1696, \dodoi{10.1093/mnras/stac480}

\bibitem[{{Ho} {et~al.}(2021){Ho}, {Goto}, {Oi}, {Kim}, {Malkan}, {Pollo},
  {Hashimoto}, {Toba}, {Kim}, {Hwang}, {Shim}, {Huang}, {Kim}, {Wang},
  {Santos}, \& {Matsuhara}}]{Ho}
{Ho}, S. C.~C., {Goto}, T., {Oi}, N., {et~al.} 2021, \mnras, 502, 140,
  \dodoi{10.1093/mnras/staa3549}

\bibitem[{{Hopfield}(1982)}]{1982PNAS...79.2554H}
{Hopfield}, J.~J. 1982, Proceedings of the National Academy of Science, 79,
  2554, \dodoi{10.1073/pnas.79.8.2554}

\bibitem[{{Huang} {et~al.}(2020){Huang}, {Matsuhara}, {Goto}, {Shim}, {Kim},
  {Malkan}, {Hashimoto}, {Hwang}, {Oi}, {Toba}, {Lee}, {Santos}, \&
  {Takagi}}]{megaprime}
{Huang}, T.-C., {Matsuhara}, H., {Goto}, T., {et~al.} 2020, \mnras, 498, 609,
  \dodoi{10.1093/mnras/staa2459}

\bibitem[{{Huang} {et~al.}(2021){Huang}, {Matsuhara}, {Goto}, {Santos}, {Ho},
  {Kim}, {Hashimoto}, {Ikeda}, {Oi}, {Malkan}, {Pearson}, {Pollo}, {Serjeant},
  {Shim}, {Miyaji}, {Hwang}, {Durkalec}, {Poliszczuk}, {Greve}, {Pearson},
  {Toba}, {Lee}, {Kim}, {Toft}, {Jeong}, \& {Enokidani}}]{huang21}
---. 2021, \mnras, 506, 6063, \dodoi{10.1093/mnras/stab2128}

\bibitem[{{Hwang} {et~al.}(2007){Hwang}, {Lee}, {Lee}, {Im}, {Kim},
  {Matsuhara}, {Wada}, {Oyabu}, {Pak}, {Chun}, {Watarai}, {Nakagawa},
  {Pearson}, {Takagi}, {Hanami}, \& {White}}]{cfht_obs1}
{Hwang}, N., {Lee}, M.~G., {Lee}, H.~M., {et~al.} 2007, \apjs, 172, 583,
  \dodoi{10.1086/519216}

\bibitem[{{Ilbert} {et~al.}(2006{\natexlab{a}}){Ilbert}, {Arnouts},
  {McCracken}, {Bolzonella}, {Bertin}, {Le F{\`e}vre}, {Mellier}, {Zamorani},
  {Pell{\`o}}, {Iovino}, {Tresse}, {Le Brun}, {Bottini}, {Garilli}, {Maccagni},
  {Picat}, {Scaramella}, {Scodeggio}, {Vettolani}, {Zanichelli}, {Adami},
  {Bardelli}, {Cappi}, {Charlot}, {Ciliegi}, {Contini}, {Cucciati}, {Foucaud},
  {Franzetti}, {Gavignaud}, {Guzzo}, {Marano}, {Marinoni}, {Mazure}, {Meneux},
  {Merighi}, {Paltani}, {Pollo}, {Pozzetti}, {Radovich}, {Zucca}, {Bondi},
  {Bongiorno}, {Busarello}, {de La Torre}, {Gregorini}, {Lamareille}, {Mathez},
  {Merluzzi}, {Ripepi}, {Rizzo}, \& {Vergani}}]{intro3}
{Ilbert}, O., {Arnouts}, S., {McCracken}, H.~J., {et~al.} 2006{\natexlab{a}},
  \aap, 457, 841, \dodoi{10.1051/0004-6361:20065138}

\bibitem[{{Ilbert} {et~al.}(2006{\natexlab{b}}){Ilbert}, {Arnouts},
  {McCracken}, {Bolzonella}, {Bertin}, {Le F{\`e}vre}, {Mellier}, {Zamorani},
  {Pell{\`o}}, {Iovino}, {Tresse}, {Le Brun}, {Bottini}, {Garilli}, {Maccagni},
  {Picat}, {Scaramella}, {Scodeggio}, {Vettolani}, {Zanichelli}, {Adami},
  {Bardelli}, {Cappi}, {Charlot}, {Ciliegi}, {Contini}, {Cucciati}, {Foucaud},
  {Franzetti}, {Gavignaud}, {Guzzo}, {Marano}, {Marinoni}, {Mazure}, {Meneux},
  {Merighi}, {Paltani}, {Pollo}, {Pozzetti}, {Radovich}, {Zucca}, {Bondi},
  {Bongiorno}, {Busarello}, {de La Torre}, {Gregorini}, {Lamareille}, {Mathez},
  {Merluzzi}, {Ripepi}, {Rizzo}, \& {Vergani}}]{Ilbert06}
---. 2006{\natexlab{b}}, \aap, 457, 841, \dodoi{10.1051/0004-6361:20065138}

\bibitem[{James {et~al.}(2021)James, Witten, Hastie, \& Tibshirani}]{ML}
James, G., Witten, D., Hastie, T., \& Tibshirani, R. 2021, An introduction to
  Statistical Learning, 2nd edn. (NY: Springer New York),
  \dodoi{10.1007/978-1-0716-1418-1}

\bibitem[{{Jansen} \& {Windhorst}(2018)}]{2018PASP..130l4001J}
{Jansen}, R.~A., \& {Windhorst}, R.~A. 2018, \pasp, 130, 124001,
  \dodoi{10.1088/1538-3873/aae476}

\bibitem[{{Jarrett} {et~al.}(2011){Jarrett}, {Cohen}, {Masci}, {Wright},
  {Stern}, {Benford}, {Blain}, {Carey}, {Cutri}, {Eisenhardt}, {Lonsdale},
  {Mainzer}, {Marsh}, {Padgett}, {Petty}, {Ressler}, {Skrutskie}, {Stanford},
  {Surace}, {Tsai}, {Wheelock}, \& {Yan}}]{wise}
{Jarrett}, T.~H., {Cohen}, M., {Masci}, F., {et~al.} 2011, \apj, 735, 112,
  \dodoi{10.1088/0004-637X/735/2/112}

\bibitem[{{Jeon} {et~al.}(2010){Jeon}, {Im}, {Ibrahimov}, {Lee}, {Lee}, \&
  {Lee}}]{maidanak_obs}
{Jeon}, Y., {Im}, M., {Ibrahimov}, M., {et~al.} 2010, \apjs, 190, 166,
  \dodoi{10.1088/0067-0049/190/1/166}

\bibitem[{{Jeon} {et~al.}(2014){Jeon}, {Im}, {Kang}, {Lee}, \&
  {Matsuhara}}]{2014ApJS..214...20J}
{Jeon}, Y., {Im}, M., {Kang}, E., {Lee}, H.~M., \& {Matsuhara}, H. 2014, \apjs,
  214, 20, \dodoi{10.1088/0067-0049/214/2/20}

\bibitem[{{Kim} {et~al.}(2021{\natexlab{a}}){Kim}, {Hwang}, {Jeong}, {Kim},
  {Burgarella}, {Goto}, {Hashimoto}, {Jo}, {Lee}, {Malkan}, {Pearson}, {Shim},
  {Toba}, {Ho}, {Santos}, {Ikeda}, {Kim}, {Miyaji}, {Matsuhara}, {Oi},
  {Takagi}, \& {Wang}}]{kimeb21}
{Kim}, E., {Hwang}, H.~S., {Jeong}, W.-S., {et~al.} 2021{\natexlab{a}}, \mnras,
  507, 3113, \dodoi{10.1093/mnras/stab2090}

\bibitem[{{Kim} {et~al.}(2018){Kim}, {Malkan}, {Oi}, {Burgarella}, {Buat},
  {Takagi}, \& {Matsuhara}}]{specz4}
{Kim}, H.~K., {Malkan}, M.~A., {Oi}, N., {et~al.} 2018, in The Cosmic Wheel and
  the Legacy of the AKARI Archive: From Galaxies and Stars to Planets and Life,
  ed. T.~{Ootsubo}, I.~{Yamamura}, K.~{Murata}, \& T.~{Onaka}

\bibitem[{{Kim} {et~al.}(2012){Kim}, {Lee}, {Matsuhara}, {Wada}, {Oyabu}, {Im},
  {Jeon}, {Kang}, {Ko}, {Lee}, {Takagi}, {Pearson}, {White}, {Jeong},
  {Serjeant}, {Nakagawa}, {Ohyama}, {Goto}, {Takeuchi}, {Pollo}, {Solarz}, \&
  {P{\c{e}}piak}}]{NEPW_survey}
{Kim}, S.~J., {Lee}, H.~M., {Matsuhara}, H., {et~al.} 2012, \aap, 548, A29,
  \dodoi{10.1051/0004-6361/201219105}

\bibitem[{{Kim} {et~al.}(2019){Kim}, {Jeong}, {Goto}, {Lee}, {Shim}, {Pearson},
  {Im}, {Matsuhara}, {Seo}, {Hashimoto}, {Kim}, {Chiang}, {Barrufet}, \&
  {Varillas}}]{kimsj19}
{Kim}, S.~J., {Jeong}, W.-S., {Goto}, T., {et~al.} 2019, \pasj, 71, 11,
  \dodoi{10.1093/pasj/psy121}

\bibitem[{{Kim} {et~al.}(2021{\natexlab{b}}){Kim}, {Oi}, {Goto}, {Ikeda}, {Ho},
  {Shim}, {Toba}, {Hwang}, {Hashimoto}, {Barrufet}, {Malkan}, {Kim}, {Huang},
  {Matsuhara}, {Miyaji}, {Pearson}, {Serjeant}, {Santos}, {Kim}, {Pollo},
  {Jeong}, {Wang}, {Momose}, \& {Takagi}}]{NEPW_catalog}
{Kim}, S.~J., {Oi}, N., {Goto}, T., {et~al.} 2021{\natexlab{b}}, \mnras, 500,
  4078, \dodoi{10.1093/mnras/staa3359}

\bibitem[{{Krumpe} {et~al.}(2015){Krumpe}, {Miyaji}, {Brunner}, {Hanami},
  {Ishigaki}, {Takagi}, {Markowitz}, {Goto}, {Malkan}, {Matsuhara}, {Pearson},
  {Ueda}, \& {Wada}}]{Krumpe2015}
{Krumpe}, M., {Miyaji}, T., {Brunner}, H., {et~al.} 2015, \mnras, 446, 911,
  \dodoi{10.1093/mnras/stu2010}

\bibitem[{{Kurtz} \& {Mink}(1998)}]{rvsao}
{Kurtz}, M.~J., \& {Mink}, D.~J. 1998, \pasp, 110, 934, \dodoi{10.1086/316207}

\bibitem[{{Laigle} {et~al.}(2016){Laigle}, {McCracken}, {Ilbert}, {Hsieh},
  {Davidzon}, {Capak}, {Hasinger}, {Silverman}, {Pichon}, {Coupon}, {Aussel},
  {Le Borgne}, {Caputi}, {Cassata}, {Chang}, {Civano}, {Dunlop}, {Fynbo},
  {Kartaltepe}, {Koekemoer}, {Le F{\`e}vre}, {Le Floc'h}, {Leauthaud}, {Lilly},
  {Lin}, {Marchesi}, {Milvang-Jensen}, {Salvato}, {Sanders}, {Scoville},
  {Smolcic}, {Stockmann}, {Taniguchi}, {Tasca}, {Toft}, {Vaccari}, \&
  {Zabl}}]{Laigle2016}
{Laigle}, C., {McCracken}, H.~J., {Ilbert}, O., {et~al.} 2016, \apjs, 224, 24,
  \dodoi{10.3847/0067-0049/224/2/24}

\bibitem[{{Lee} {et~al.}(2009){Lee}, {Kim}, {Im}, {Matsuhara}, {Oyabu}, {Wada},
  {Nakagawa}, {Ko}, {Shim}, {Lee}, {Hwang}, {Takagi}, \&
  {Pearson}}]{2009PASJ...61..375L}
{Lee}, H.~M., {Kim}, S.~J., {Im}, M., {et~al.} 2009, \pasj, 61, 375,
  \dodoi{10.1093/pasj/61.2.375}

\bibitem[{{Lee} \& {Shin}(2021)}]{2021AJ....162..297L}
{Lee}, J., \& {Shin}, M.-S. 2021, \aj, 162, 297,
  \dodoi{10.3847/1538-3881/ac2e96}

\bibitem[{{Leistedt} \& {Hogg}(2017)}]{2017ApJ...838....5L}
{Leistedt}, B., \& {Hogg}, D.~W. 2017, \apj, 838, 5,
  \dodoi{10.3847/1538-4357/aa6332}

\bibitem[{{Luo} {et~al.}(2024){Luo}, {Li}, {Lu}, {Chen}, {Fu}, {Zhang}, {Xiao},
  {Du}, {Gong}, {Shu}, {Ma}, {Meng}, {Zhou}, \& {Fan}}]{Luo2024}
{Luo}, Z., {Li}, Y., {Lu}, J., {et~al.} 2024, \mnras,
  \dodoi{10.1093/mnras/stae2446}

\bibitem[{{Matsuhara} {et~al.}(2006){Matsuhara}, {Wada}, {Matsuura},
  {Nakagawa}, {Kawada}, {Ohyama}, {Pearson}, {Oyabu}, {Takagi}, {Serjeant},
  {White}, {Hanami}, {Watarai}, {Takeuchi}, {Kodama}, {Arimoto}, {Okamura},
  {Lee}, {Pak}, {Im}, {Lee}, {Kim}, {Jeong}, {Imai}, {Fujishiro}, {Shirahata},
  {Suzuki}, {Ihara}, \& {Sakon}}]{NEP}
{Matsuhara}, H., {Wada}, T., {Matsuura}, S., {et~al.} 2006, \pasj, 58, 673,
  \dodoi{10.1093/pasj/58.4.673}

\bibitem[{{Miyaji} {et~al.}(2024){Miyaji}, {Bravo-Navarro}, {D{\'\i}az Tello},
  {Krumpe}, {Herrera-Endoqui}, {Ikeda}, {Takagi}, {Oi}, {Shogaki}, {Matsuura},
  {Kim}, {Malkan}, {Hwang}, {Kim}, {Ishigaki}, {Hanami}, {Kim}, {Ohyama},
  {Goto}, \& {Matsuhara}}]{Miyaji24}
{Miyaji}, T., {Bravo-Navarro}, B.~A., {D{\'\i}az Tello}, J., {et~al.} 2024,
  \aap, 689, A83, \dodoi{10.1051/0004-6361/202450453}

\bibitem[{{Murakami} {et~al.}(2007){Murakami}, {Baba}, {Barthel}, {Clements},
  {Cohen}, {Doi}, {Enya}, {Figueredo}, {Fujishiro}, {Fujiwara}, {Fujiwara},
  {Garcia-Lario}, {Goto}, {Hasegawa}, {Hibi}, {Hirao}, {Hiromoto}, {Hong},
  {Imai}, {Ishigaki}, {Ishiguro}, {Ishihara}, {Ita}, {Jeong}, {Jeong},
  {Kaneda}, {Kataza}, {Kawada}, {Kawai}, {Kawamura}, {Kessler}, {Kester},
  {Kii}, {Kim}, {Kim}, {Kobayashi}, {Koo}, {Kwon}, {Lee}, {Lorente}, {Makiuti},
  {Matsuhara}, {Matsumoto}, {Matsuo}, {Matsuura}, {M{\"U}ller}, {Murakami},
  {Nagata}, {Nakagawa}, {Naoi}, {Narita}, {Noda}, {Oh}, {Ohnishi}, {Ohyama},
  {Okada}, {Okuda}, {Oliver}, {Onaka}, {Ootsubo}, {Oyabu}, {Pak}, {Park},
  {Pearson}, {Rowan-Robinson}, {Saito}, {Sakon}, {Salama}, {Sato}, {Savage},
  {Serjeant}, {Shibai}, {Shirahata}, {Sohn}, {Suzuki}, {Takagi}, {Takahashi},
  {Tanab{\'E}}, {Takeuchi}, {Takita}, {Thomson}, {Uemizu}, {Ueno}, {Usui},
  {Verdugo}, {Wada}, {Wang}, {Watabe}, {Watarai}, {White}, {Yamamura},
  {Yamauchi}, \& {Yasuda}}]{2007PASJ...59S.369M}
{Murakami}, H., {Baba}, H., {Barthel}, P., {et~al.} 2007, \pasj, 59, S369,
  \dodoi{10.1093/pasj/59.sp2.S369}

\bibitem[{{Nayyeri} {et~al.}(2018){Nayyeri}, {Ghotbi}, {Cooray}, {Bock},
  {Clements}, {Im}, {Kim}, {Korngut}, {Lanz}, {Lee}, {Lee}, {Malkan},
  {Matsuhara}, {Matsumoto}, {Matsuura}, {Nam}, {Pearson}, {Serjeant}, {Smidt},
  {Tsumura}, {Wada}, \& {Zemcov}}]{spitzer}
{Nayyeri}, H., {Ghotbi}, N., {Cooray}, A., {et~al.} 2018, \apjs, 234, 38,
  \dodoi{10.3847/1538-4365/aaa07e}

\bibitem[{{Ohyama} {et~al.}(2018){Ohyama}, {Wada}, {Matsuhara}, {Takagi},
  {Malkan}, {Goto}, {Egami}, {Lee}, {Im}, {Kim}, {Pearson}, {Inami}, {Oyabu},
  {Usui}, {Burgarella}, {Mazyed}, {Imanishi}, {Jeong}, {Miyaji}, {D{\'\i}az
  Tello}, {Nakagawa}, {Serjeant}, {Takeuchi}, {Toba}, {White}, {Hanami}, \&
  {Ishigaki}}]{specz7}
{Ohyama}, Y., {Wada}, T., {Matsuhara}, H., {et~al.} 2018, \aap, 618, A101,
  \dodoi{10.1051/0004-6361/201731470}

\bibitem[{Oi {et~al.}(2017)Oi, Goto, Malkan, Pearson, \& Matsuhara}]{specz6}
Oi, N., Goto, T., Malkan, M., Pearson, C., \& Matsuhara, H. 2017, Publications
  of the Astronomical Society of Japan, 69, \dodoi{10.1093/pasj/psx053}

\bibitem[{{Oi} {et~al.}(2014){Oi}, {Matsuhara}, {Murata}, {Goto}, {Wada},
  {Takagi}, {Ohyama}, {Malkan}, {Im}, {Shim}, {Serjeant}, \&
  {Pearson}}]{cfht_obs2}
{Oi}, N., {Matsuhara}, H., {Murata}, K., {et~al.} 2014, \aap, 566, A60,
  \dodoi{10.1051/0004-6361/201322561}

\bibitem[{{Oi} {et~al.}(2021){Oi}, {Goto}, {Matsuhara}, {Utsumi}, {Momose},
  {Toba}, {Malkan}, {Takagi}, {Huang}, {Kim}, \& {Ohyama}}]{deep_hsc2}
{Oi}, N., {Goto}, T., {Matsuhara}, H., {et~al.} 2021, \mnras, 500, 5024,
  \dodoi{10.1093/mnras/staa3080}

\bibitem[{{Pasquet} {et~al.}(2019){Pasquet}, {Bertin}, {Treyer}, {Arnouts}, \&
  {Fouchez}}]{Pasquet2019}
{Pasquet}, J., {Bertin}, E., {Treyer}, M., {Arnouts}, S., \& {Fouchez}, D.
  2019, \aap, 621, A26, \dodoi{10.1051/0004-6361/201833617}

\bibitem[{Pedregosa {et~al.}(2011)Pedregosa, Varoquaux, Gramfort, Michel,
  Thirion, Grisel, Blondel, Prettenhofer, Weiss, Dubourg, Vanderplas, Passos,
  Cournapeau, Brucher, Perrot, \& Duchesnay}]{scikit-learn}
Pedregosa, F., Varoquaux, G., Gramfort, A., {et~al.} 2011, Journal of Machine
  Learning Research, 12, 2825

\bibitem[{{Poliszczuk} {et~al.}(2021){Poliszczuk}, {Pollo}, {Ma{\l}ek},
  {Durkalec}, {Pearson}, {Goto}, {Kim}, {Malkan}, {Oi}, {Ho}, {Shim},
  {Pearson}, {Hwang}, {Toba}, \& {Kim}}]{Poliszczuk21}
{Poliszczuk}, A., {Pollo}, A., {Ma{\l}ek}, K., {et~al.} 2021, \aap, 651, A108,
  \dodoi{10.1051/0004-6361/202040219}

\bibitem[{{Puschell} {et~al.}(1982){Puschell}, {Owen}, \&
  {Laing}}]{Puschell1982}
{Puschell}, J.~J., {Owen}, F.~N., \& {Laing}, R.~A. 1982, \apjl, 257, L57,
  \dodoi{10.1086/183808}

\bibitem[{{Rasmussen} \& {Williams}(2006)}]{2006gpml.book.....R}
{Rasmussen}, C.~E., \& {Williams}, C. K.~I. 2006, {Gaussian Processes for
  Machine Learning}

\bibitem[{{Salvato} {et~al.}(2019){Salvato}, {Ilbert}, \& {Hoyle}}]{photoz}
{Salvato}, M., {Ilbert}, O., \& {Hoyle}, B. 2019, Nature Astronomy, 3, 212,
  \dodoi{10.1038/s41550-018-0478-0}

\bibitem[{{Santos} {et~al.}(2021){Santos}, {Goto}, {Kim}, {Wang}, {Ho},
  {Hashimoto}, {Huang}, {Lu}, {On}, {Wong}, {Hsiao}, {Pollo}, {Malkan},
  {Miyaji}, {Toba}, {Kilerci-Eser}, {Ma{\l}ek}, {Hwang}, {Jeong}, {Shim},
  {Pearson}, {Poliszczuk}, \& {Chen}}]{santos21}
{Santos}, D. J.~D., {Goto}, T., {Kim}, S.~J., {et~al.} 2021, \mnras, 507, 3070,
  \dodoi{10.1093/mnras/stab2352}

\bibitem[{{Schuldt} {et~al.}(2021){Schuldt}, {Suyu}, {Ca{\~n}ameras},
  {Taubenberger}, {Meinhardt}, {Leal-Taix{\'e}}, \& {Hsieh}}]{Schuldt2021}
{Schuldt}, S., {Suyu}, S.~H., {Ca{\~n}ameras}, R., {et~al.} 2021, \aap, 651,
  A55, \dodoi{10.1051/0004-6361/202039945}

\bibitem[{{Shim} {et~al.}(2013){Shim}, {Im}, {Ko}, {Jeon}, {Karouzos}, {Kim},
  {Lee}, {Papovich}, {Willmer}, \& {Weiner}}]{specz1}
{Shim}, H., {Im}, M., {Ko}, J., {et~al.} 2013, \apjs, 207, 37,
  \dodoi{10.1088/0067-0049/207/2/37}

\bibitem[{{Shim} {et~al.}(2023){Shim}, {Hwang}, {Jeong}, {Toba}, {Kim}, {Kim},
  {Song}, {Hashimoto}, {Nakagawa}, {Nanni}, {Pearson}, \&
  {Takagi}}]{2023AJ....165...31S}
{Shim}, H., {Hwang}, H.~S., {Jeong}, W.-S., {et~al.} 2023, \aj, 165, 31,
  \dodoi{10.3847/1538-3881/aca09c}

\bibitem[{{Shogaki} {et~al.}(2018){Shogaki}, {Matsuura}, {Oi}, {Goto},
  {Matsuhara}, {Murata}, {Takagi}, Takuya~{Otsuka}, {Malkan}, \&
  {Churei}}]{specz3}
{Shogaki}, A., {Matsuura}, S., {Oi}, N., {et~al.} 2018, in The Cosmic Wheel and
  the Legacy of the AKARI Archive: From Galaxies and Stars to Planets and Life,
  ed. T.~{Ootsubo}, I.~{Yamamura}, K.~{Murata}, \& T.~{Onaka}

\bibitem[{{Sohn} {et~al.}(2023){Sohn}, {Geller}, {Hwang}, {Fabricant},
  {Utsumi}, \& {Damjanov}}]{hectomap}
{Sohn}, J., {Geller}, M.~J., {Hwang}, H.~S., {et~al.} 2023, \apj, 945, 94,
  \dodoi{10.3847/1538-4357/acb925}

\bibitem[{{Soo} {et~al.}(2021){Soo}, {Joachimi}, {Eriksen}, {Siudek},
  {Alarcon}, {Cabayol}, {Carretero}, {Casas}, {Castander}, {Fern{\'a}ndez},
  {Garc{\'\i}a-Bellido}, {Gaztanaga}, {Hildebrandt}, {Hoekstra}, {Miquel},
  {Padilla}, {S{\'a}nchez}, {Serrano}, \& {Tallada-Cresp{\'\i}}}]{Soo2021}
{Soo}, J. Y.~H., {Joachimi}, B., {Eriksen}, M., {et~al.} 2021, \mnras, 503,
  4118, \dodoi{10.1093/mnras/stab711}

\bibitem[{{Strauss} \& {Willick}(1995)}]{intro2}
{Strauss}, M.~A., \& {Willick}, J.~A. 1995, \physrep, 261, 271,
  \dodoi{10.1016/0370-1573(95)00013-7}

\bibitem[{{Takagi} {et~al.}(2010){Takagi}, {Ohyama}, {Goto}, {Matsuhara},
  {Oyabu}, {Wada}, {Pearson}, {Lee}, {Im}, {Lee}, {Shim}, {Hanami}, {Ishigaki},
  {Imai}, {White}, {Serjeant}, \& {Malkan}}]{specz2}
{Takagi}, T., {Ohyama}, Y., {Goto}, T., {et~al.} 2010, \aap, 514, A5,
  \dodoi{10.1051/0004-6361/200913466}

\bibitem[{{Tanaka}(2015)}]{Tanaka2015}
{Tanaka}, M. 2015, \apj, 801, 20, \dodoi{10.1088/0004-637X/801/1/20}

\bibitem[{{Toba} {et~al.}(2020){Toba}, {Goto}, {Oi}, {Wang}, {Kim}, {Ho},
  {Burgarella}, {Hashimoto}, {Hsieh}, {Huang}, {Hwang}, {Ikeda}, {Kim}, {Kim},
  {Lee}, {Malkan}, {Matsuhara}, {Miyaji}, {Momose}, {Ohyama}, {Oyabu},
  {Pearson}, {Santos}, {Shim}, {Takagi}, {Ueda}, {Utsumi}, \& {Wada}}]{toba20}
{Toba}, Y., {Goto}, T., {Oi}, N., {et~al.} 2020, \apj, 899, 35,
  \dodoi{10.3847/1538-4357/ab9cb7}

\bibitem[{{Tonry} \& {Davis}(1979)}]{td79}
{Tonry}, J., \& {Davis}, M. 1979, \aj, 84, 1511, \dodoi{10.1086/112569}

\bibitem[{{Vanderplas} {et~al.}(2012){Vanderplas}, {Connolly}, {Ivezi{\'c}}, \&
  {Gray}}]{astroML}
{Vanderplas}, J., {Connolly}, A., {Ivezi{\'c}}, {\v Z}., \& {Gray}, A. 2012, in
  Conference on Intelligent Data Understanding (CIDU), 47 --54,
  \dodoi{10.1109/CIDU.2012.6382200}

\bibitem[{{Wang} {et~al.}(2020){Wang}, {Goto}, {Kim}, {Hashimoto},
  {Burgarella}, {Toba}, {Shim}, {Miyaji}, {Hwang}, {Jeong}, {Kim}, {Ikeda},
  {Pearson}, {Malkan}, {Oi}, {Santos}, {Ma{\l}ek}, {Pollo}, {Ho}, {Matsuhara},
  {On}, {Kim}, {Hsiao}, \& {Huang}}]{wang20}
{Wang}, T.-W., {Goto}, T., {Kim}, S.~J., {et~al.} 2020, \mnras, 499, 4068,
  \dodoi{10.1093/mnras/staa2988}

\bibitem[{{Werner} {et~al.}(2004){Werner}, {Roellig}, {Low}, {Rieke}, {Rieke},
  {Hoffmann}, {Young}, {Houck}, {Fazio}, {Hora}, {Gehrz}, {Soifer}, {Helou},
  {Keene}, {Eisenhardt}, {Gallagher}, {Gautier}, {Irace}, {Lawrence},
  {Simmons}, {Wright}, {Jura}, {Cruikshank}, \& {Brandl}}]{2004AAS...204.3301W}
{Werner}, M., {Roellig}, T.~L., {Low}, F.~J., {et~al.} 2004, in American
  Astronomical Society Meeting Abstracts, Vol. 204, American Astronomical
  Society Meeting Abstracts \#204, 33.01

\bibitem[{{Wright} {et~al.}(2010){Wright}, {Eisenhardt}, {Mainzer}, {Ressler},
  {Cutri}, {Jarrett}, {Kirkpatrick}, {Padgett}, {McMillan}, {Skrutskie},
  {Stanford}, {Cohen}, {Walker}, {Mather}, {Leisawitz}, {Gautier}, {McLean},
  {Benford}, {Lonsdale}, {Blain}, {Mendez}, {Irace}, {Duval}, {Liu}, {Royer},
  {Heinrichsen}, {Howard}, {Shannon}, {Kendall}, {Walsh}, {Larsen}, {Cardon},
  {Schick}, {Schwalm}, {Abid}, {Fabinsky}, {Naes}, \&
  {Tsai}}]{2010AJ....140.1868W}
{Wright}, E.~L., {Eisenhardt}, P. R.~M., {Mainzer}, A.~K., {et~al.} 2010, \aj,
  140, 1868, \dodoi{10.1088/0004-6256/140/6/1868}

\end{thebibliography}
\bibliographystyle{aasjournal}

\end{document}